\documentclass[twocolumn,journal]{IEEEtran}
\usepackage[T1]{fontenc}
\usepackage{xcolor}
\usepackage{amsmath}
\usepackage{amssymb}
\usepackage{graphicx}
\PassOptionsToPackage{normalem}{ulem}
\usepackage{ulem}
\usepackage[keeplastbox]{flushend}

\makeatletter

\providecommand{\tabularnewline}{\\}

\usepackage[caption=false,font=footnotesize]{subfig}
\usepackage{amsmath}

\@ifundefined{showcaptionsetup}{}{%
 \PassOptionsToPackage{caption=false}{subfig}}
\usepackage{subfig}
\makeatother

\begin{document}

\title{Analysis of High Frequency Impedance Measurement Techniques for Power
Line Network Sensing}

\author{Federico Passerini and~Andrea
M. Tonello\thanks{The authors are with the Networked and
Embedded Systems Institute, University of Klagenfurt, Klagenfurt,
Austria, e-mail: \{federico.passerini, andrea.tonello\}@aau.at.
\newline \noindent
A version of this article has been accepted for publication on IEEE Sensors Journal, Special Issue on ``Smart Sensors for Smart Grids and Smart Cities''.

\noindent IEEE Copyright Notice: \textsf{\copyright} 2017 IEEE. Personal use of this material is permitted. Permission from IEEE must be obtained for all other uses, in any current or future media, including reprinting/republishing this material for advertising or promotional purposes, creating new collective works, for resale or redistribution to servers or lists, or reuse of any copyrighted component of this work in other works.}}
\maketitle
\begin{abstract}
A major aspect in power line distribution networks is
the constant monitoring of the network properties. With the advent
of the smart grid concept, distributed monitoring has started complementing
the information of the central stations. In this context, power line
communications modems deployed throughout the network provide a tool
to monitor high frequency components of the signals traveling through
a power line network. We propose therefore to use them not only as
communication devices but also as network sensors. Besides classical
voltage measurements, these sensors can be designed to monitor high
frequency impedances, which provide useful information about the power
line network, as for instance status of the topology, cable degradation
and occurrence of faults. In this article, we provide a technical
analysis of different voltage and impedance measurement techniques
that can be integrated into power line modems. We assess the accuracy
of the techniques under analysis in the presence of network noise
and we discuss the statistical characteristics of the measurement
noise. We finally compare the performances of the examined techniques
when applied to the fault detection problem in distribution networks,
in order to establish which technique gives more accurate results. \end{abstract}


\IEEEpeerreviewmaketitle{}

\section{Introduction}

\IEEEPARstart{T}{he} concept of Smart Grids (SG) refers to the faculty
of power grids to operate at the maximum efficiency with the minimum
cost in a self-organized fashion, or with minimum human intervention.
Within any SG, a huge set of information is sensed and shared throughout
the network, then processed by optimization algorithms that subsequently
control the network parameters, usually aiming at the maximum efficiency/cost
ratio \cite{bush2014smartgrid}. The first element of this cumbersome
chain of operations comprises a network of sensors, whose precision
and reliability gives the most fundamental contribution to the overall
performance of the SG. The topic of this paper is the study
of the accuracy of the measurement performed by the network sensors,
and in particular by impedance and voltage sensors. 

Most of the network sensors are branched to the electrical grid and
measure either voltages or currents, or both. Such quantities are
then processed and analyzed to monitor different aspects of the network
\cite{6099519}. Measurements at the mains frequency are usually performed
in transmission networks by phasor measurement units (PMU) in order
to monitor the energy consumption and the load balance, but also to
detect and localize possible faults \cite{6814841}. Other techniques
used for the same purposes include measurements at the mains harmonics
up to few kHz, using either pulsed, sinusoidal or wavelet test signals
\cite{faultreview}. Fault detection has also beed performed by means
of surface waves \cite{6295639}. All these sensors can be used also
in medium voltage (MV) or low voltage (LV) distribution networks for
the same purposes. However, distribution networks are characterized
by specific issues: the complex and often changing topological structure
of these networks has to be constantly monitored \cite{7779155};
high impedance faults (HIF) are common and often cause undetectable
damage to the network, which on a medium to long run can cause a complete
system failure \cite{faultreview}; the power cables are often insulated
and buried underground, where they are subject to aging due to water
treeing, oxidation and other causes \cite{Tawancy2016}. To tackle
these issues, different methods have been proposed that make use of
high frequency measurements ranging from few kHz up to some MHz. In
particular, frequency domain reflectometry (FDR) is emerging as a
promising technique, thanks to the simplicity of processing the signal to 
obtain accurate results. FDR consists of transmitting a broadband or swept
voltage signal into the network and correlating the signal reflected
back to the transmitted one. This procedure can provide valuable information
about the topology of the network \cite{ahmed2012topology2}, the
presence and location of faults \cite{zhangphd} and also the electrical
characteristics of the power cables. Similarly, dividing the reflected
signal by the transmitted signal provides a measure of the reflection
coefficient $\rho$, which allows to retrieve the same information
as FDR does \cite{neusphd}. Performing measurements in frequency
domain enables the use of signal processing techniques like windowing,
zero padding, warping and compensation, which highly improve the quality
of the extracted information. More recently, other high frequency
techniques have been proposed that make use of impedance measurements.
In particular, \cite{7476292} shows that impedance measurement techniques
can provide as much information as FDR techniques does for the purpose
of fault detection.. In addition, the use of impedance measurements
enables the development of new methods for topology estimation \cite{io}
, resulting in impedance being a more informative
 quantity than the signal trace measured with the FDR.

Given the growing interest on these measurement techniques, the aim
of this paper is to shed new light on their applicability and on their
accuracy. In particular, given the SG context and the need of smart
sensors, we propose to integrate either impedance, FDR or $\rho$
measurements in power line modems (PLM), which are largely deployed
in distribution networks to allow power line communications (PLC)
\cite{5768099}. Therefore, PLM will also act also as network sensors.

This paper considers three known impedance measurement methods along
with one known method to measure $\rho$ or the FDR trace. All these
methods are herein revisited with the aim of easy integration within
PLM. Consequently, the methods are compared based on typical PLM architectures
that operate using different PLC standards, taking into account typical
hardware noise sources and giving emphasis to the effects of the medium
noise. In fact, the presence of high noise is a major concern in PLC
\cite{5764408} that can act as a severe hurdle to both communication
and measurement devices. The effect of the noise is both computed analytically and analyzed
via simulations, the prerequisites under which the noise can be considered
additive Gaussian are assessed, and the accuracy of these measurements
is evaluated in different conditions. Moreover, we compare the accuracy
of impedance measurements with respect to FDR and reflection coefficient
measurements in order to assess which method is the most reliable
for network sensing. To this aim, we apply all the presented techniques
to the fault detection problem and analyze which technique can detect
the fault more accurately.

The rest of the paper is organized as follows. A brief review of the
information that can be harvested form single-ended network measurements
is presented in Section \ref{sec:Transmission-line-theory}. Section
\ref{sec:Impedance-measurement-techniques} presents the considered
impedance measurement techniques and the analytical computation of
the effect of noise. The same is done for the FDR and $\rho$ measurements
in Section \ref{sec:FDR-measurement-technique}. Section \ref{sec:Considerations-on-the}
presents the PLC features that are implemented in the simulation for
the comparison of the techniques, whose results are then exposed and
commented in Section \ref{sec:Comparison-of-the}. Conclusions follow
in Section \ref{sec:Conclusions}. As a remark, all the physical quantities
and equations that we present in Sections \ref{sec:Transmission-line-theory},
\ref{sec:Impedance-measurement-techniques} and \ref{sec:FDR-measurement-technique}
are frequency dependent, but we chose not to explicitly report this
dependency to ease the notation.

\section{Single-ended network sensing\label{sec:Transmission-line-theory}}

In this section, we present the main types of high-frequency measurements
that can be performed from a single end and we explain the concepts
derived from the transmission line (TL) theory \cite{paul2008analysis}
that allow to analyze the measurements for network sensing. 

Considering an impedance measurement device plugged to a two-conductor
TL, the power line input impedance $Z_{PL}$ is defined as
\begin{equation}
Z_{PL}=Z_{C}\frac{1+\rho}{1-\rho}=Z_{C}\frac{1+\rho_{L}e^{-2\Gamma x}}{1-\rho_{L}e^{-2\Gamma x}}\label{eq:z}
\end{equation}
where $Z_{C}$ is the characteristic impedance of the TL, $\Gamma$
is its propagation constant, $x$ is the distance from the load. $\rho$
is the reflection coefficient defined as
\begin{equation}
\rho=\frac{Z_{PL}-Z_{C}}{Z_{PL}+Z_{C}}=\frac{V_{RX}}{V_{TX}}\label{eq:rho-1}
\end{equation}
where $V_{TX}$ is the forward traveling voltage wave, i.e. the transmitted
signal, and $V_{RX}$ is the backward traveling voltage wave, i.e.
the received signal at the same end. $\rho_{L}$ is the load reflection
coefficient defined as
\begin{equation}
\rho_{L}=\frac{Z_{L}-Z_{C}}{Z_{L}+Z_{C}}
\end{equation}
where $Z_{L}$ is the load impedance. The load can be either an appliance/device,
a transformer, or simply another TL that bridges the considered TL
to the rest of the network. Finally, the so called FDR trace is defined
as
\begin{equation}
T=V_{RX}V_{TX}^{*}=\rho\left|V_{TX}\right|^{2},\label{eq:trace}
\end{equation}
which is the Fourier transform of the correlation of the transmitted
and received signals in time domain \cite{thesisTorre}. If we now
compare \eqref{eq:z}, \eqref{eq:rho-1} and \eqref{eq:trace}, we
see that these formulas can all be derived from each other with a
transformation and the application of a scaling factor. It is more
important though, to point out that measuring one of these three physical
quantities gives the same amount of information about the TL. This
is particularly true for the estimation of the line length $x$. We
showed in \cite{7897102} that taking the inverse Fourier transform
(IFT) of $Z$ measured over a sufficiently wide range of frequencies
results in a series of equidistant peaks whose inter-peak distance
is a good estimate of $x$. When considering a complex network made
of a series of branches and interconnections, a similar series of
peaks is produced for every branch, resulting at the measurement node
in an aggregate series of peaks. Similar works have been published
that make use of $\rho$ \cite{neusphd} or $T$ \cite{ahmed2012topology2}
measurements. These papers show different techniques to preprocess
the measured trace in frequency domain and analyze its IFT to estimate
the vector $\mathbf{x}$ of all the distance between the discontinuities.
The knowledge of $\mathbf{x}$ allows then the reconstruction of the
network topology using different algorithms. Similar peak analysis
based techniques have also been developed to detect and localize the
presence of electrical faults in the network \cite{7897102}.

Since $Z_{PL}$, $\rho$ and $T$ are inter-related, the choice of
one of these quantities for the network sensing resides in the measurement
technique to be deployed. In fact, in order to measure $Z_{PL}$ both
voltage and current sensing are required, while measuring $\rho$
requires a double voltage sensing. Although measuring $\rho$ is equivalent
to measuring $T$ when no noise is present, except for a scaling factor,
this last measurement has been preferred in the literature because
of its simple analog implementation. In fact, $T$ can be measured
by multiplying $V_{TX}$ and $V_{RX}$ with an analog multiplier and
passing the result through a low pass filter. $\rho$ would require
an analog signal divider instead, which is usually far more expensive.

\section{Impedance measurement techniques\label{sec:Impedance-measurement-techniques}}

Many methods have been devised to measure the impedance of electric
and electronic circuits \cite{callegaro}. The most common methods
range from current and voltage measurements to balancing or resonating
networks, including also network analysis. However, not all these
methods can be easily implemented in a PLC modem for online operation,
either because they would require too complex circuitry (balancing
bridges) or because they are intrinsically narrow band (resonating
networks). 

Conversely, easy integration in a PLM front-end can be performed for
the following techniques: Auto Balancing Bridge (PLM-ABB), current-voltage
 (PLM-IV), and Vector Network Analyzer (PLM-VNA). We therefore
consider a practical implementation where the PLM transmitter is used
as the generator of the test signal and the PLM receiver is used as
voltmeter. This implies that both the transmitter and the receiver
of a given modem are active at the same time.

We are not going to discuss implementation strategies and circuit-related
issues of the presented solutions, since these topics have been already
tackled in more specialized literature \cite{callegaro,impedancemeas}.
In this section instead, we explain at system level their operating
principles and investigate how the noise affects the measurements.
In particular, we concentrate on the effect of the electrical noise
coming from the power line channel, since it is the only source of
noise that affects the measurement independently of the modem manufacturer.
Hence, the results we are going to show have general validity. We
also consider the noise due to the digital-to-analog converter (DAC)
and line driver at the transmitter, and the noise due to the analog-to-digital
converter (ADC) at the receiver, since the resolution of DACs and
ADCs might heavily influence the accuracy of the measurements. Other
sources of noise, due to the measurement circuit, depend on the specific
technology adopted and have to be treated case-by-case.

\subsection{Measurement equations\label{sub:Measurement-equations}}

In the following, the PLM front-end transmitter is modeled with its
Thevenin equivalent with ideal source voltage $V_{S}$ and real source
impedance $R_{S}$; the power line network is modeled with its input
impedance $Z_{PL}$, which is in general complex; two different voltage
measures, $V_{a}$ and $V_{b}$, are performed with ideal volt meters.
Usually the two voltage measures are performed in series using the
same volt meter, to keep the measurement error low.
As for this subsection, any kind of noise is neglected.

\begin{figure}[tb]
\centering{}\subfloat[PLM-VNA\label{fig:Reflectometer-(VNA-or}]{\centering{}\includegraphics[width=0.7\columnwidth]{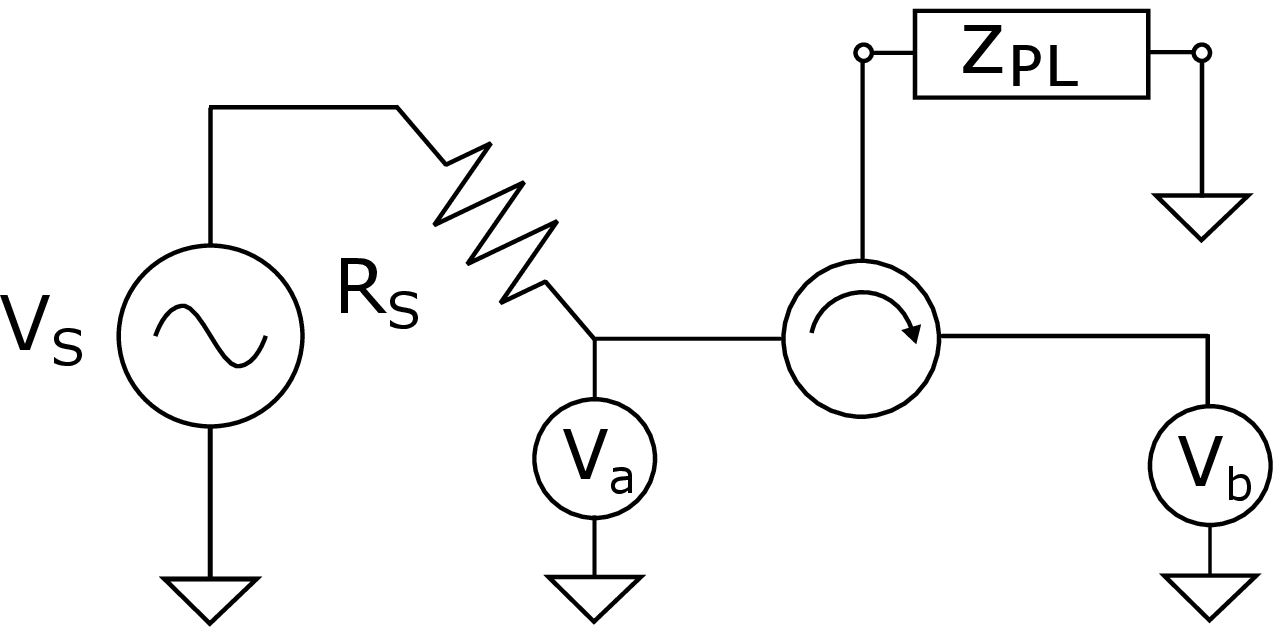}}\hfill{}\subfloat[PLM-IV\label{fig:RF-I-V-1}]{\centering{}\includegraphics[width=0.8\columnwidth]{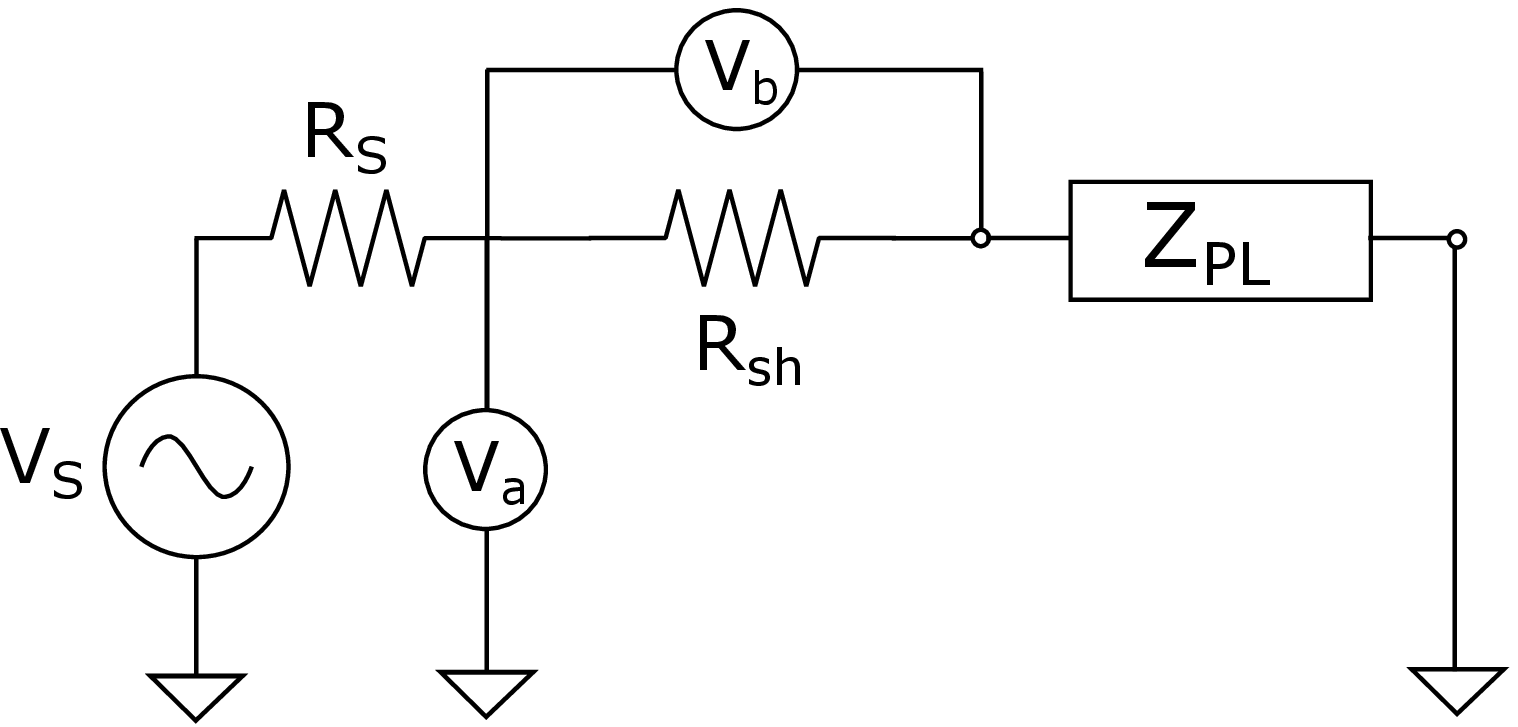}}\hfill{}\subfloat[PLM-ABB\label{fig:ABB}]{\centering{}\includegraphics[width=0.8\columnwidth]{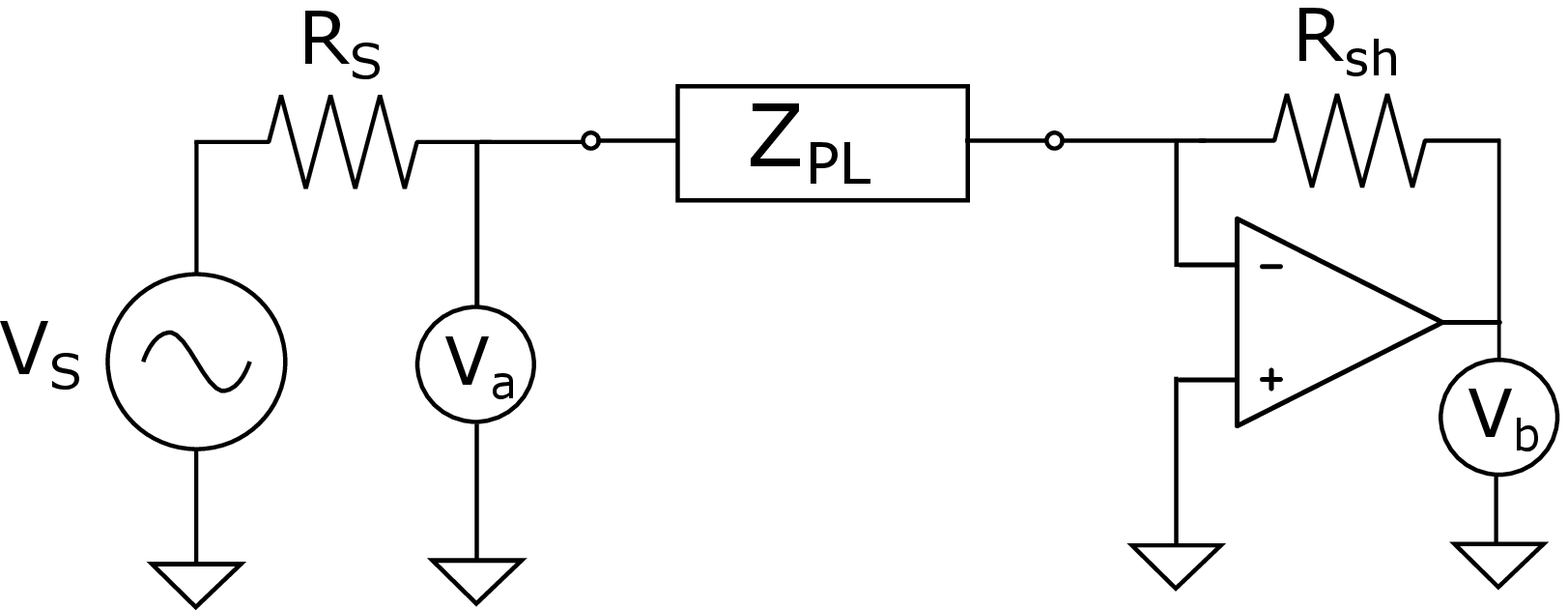}}\caption{Basic circuits of the selected impedance measurement methods.}
\end{figure}

Let us now consider the PLM-VNA architecture depicted\footnote{The depicted circuits are single-ended. Differential implementations
are also possible.} in Fig. \ref{fig:Reflectometer-(VNA-or}. The transmitter, the channel
and the receiver are connected by means of a circulator or hybrid
coupler \cite{7476291}, which also enables full-duplex communication
\cite{7505646}. The main property of the circulator is that it reflects
part of the transmitted signal to the receiver scaled by the reflection
coefficient 
\begin{equation}
\rho=\frac{Z_{PL}-Z_{o\_circ}}{Z_{PL}+Z_{o\_circ}},\label{eq:rho}
\end{equation}
 where $Z_{o\_circ}$ is the output impedance of the circulator at
the channel port. Assuming an ideal circulator, we have $V_{b}=\rho V_{a}$,
where $V_{a}$ and $V_{b}$ are the voltages measured at the transmitter
and receiver ports of the circulator, and they corespond to $V_{RX}$ and
$V_{TX}$ of the previous Section respectively. 

\begin{figure}[tb]
\centering{}\includegraphics[width=0.9\columnwidth]{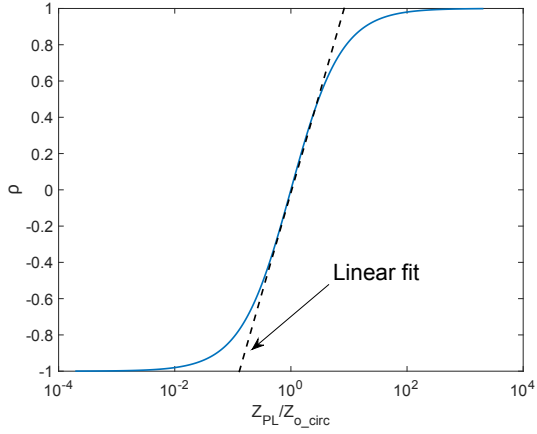}\caption{Relation between $\rho$ and $Z_{PL}$ according to \eqref{eq:rho}.\label{fig:Relation-between-}}
\end{figure}

The main purpose of a VNA is to measure $\rho$, according to the
theory of the scattering parameters \cite{Pozar}. The channel impedance
can be also derived from the measured voltages as 
\begin{equation}
Z_{PL}=Z_{o\_circ}\frac{1+\rho}{1-\rho}=Z_{o\_circ}\frac{V_{a}+V_{b}}{V_{a}-V_{b}}.\label{eq:zpln_vna}
\end{equation}
Modern VNAs can span a very wide range of frequencies, from almost
DC to several GHz, using different techniques to implement the hybrid
coupler: resistive bridges or active circuits for measurements up
to 100 MHz and strip-line couplers for higher frequencies. On the
counter side, their sensitivity falls off for impedances whose value
is far from $Z_{o\_circ}$. In fact, when $0.1<|Z_{PL}/Z_{o\_circ}|<10$
(see Fig. \ref{fig:Relation-between-}), $\rho$ almost linearly increases
with $Z_{PL}$ and so does $V_{b}$. Outside this region the slope
of $\rho$ decreases causing minor variations of $V_{b}$ due to changes
in $Z_{PL}$. This results in a deterioration of the impedance measurement
accuracy. 

The PLM-IV and PLM-ABB techniques, depicted in Fig. \ref{fig:RF-I-V-1}
and \ref{fig:ABB} respectively, do not suffer the sensitivity problem
of the PLM-VNA since they rely on the Ohm's law, which linearly relates
the ratio of the measured voltage and current to the unknown impedance.
The PLM-IV technique involves the measurement of the voltage $V_{b}$
at the edges of a shunt resistor, whose value $R_{sh}$ is known,
in series with the unknown $Z_{PL}$. This measurement provides the
value of the current flowing through $Z_{PL}$ while the voltage across
$Z_{PL}$ is given by $V_{a}$, which can be measured on either side
of the shunt resistor. In general, $V_{a}$ measured on the transmitter
side gives better results for high values of $\left|Z_{PL}\right|$,
while $V_{a}$ measured on the network side gives better results for
low values of $\left|Z_{PL}\right|$ \cite{callegaro}. Moreover,
in practice $R_{sh}$ can be replaced with a low loss transformer
that senses the current thanks to the Faraday's law. Referring to
Fig. \ref{fig:RF-I-V-1} for the following considerations, $Z_{PL}$
can be derived after simple computations as 
\begin{equation}
Z_{PL}=R_{sh}\left(\frac{V_{a}}{V_{b}}-1\right).\label{eq:zpln_rfiv}
\end{equation}

Finally, the PLM-ABB takes its name from the fact that the current
$I_{Z_{PL}}$ flowing through $Z_{PL}$ is automatically balanced
to the current $I_{R_{sh}}$ flowing through the sensing resistor
$R_{sh}$ by means of an I-V converter. The I-V converter can be a
grounded operational amplifier or a more complex integrated circuit.
The virtual ground provided by the I-V converter results in a null
input impedance. Hence, the current measurement is not affected by
the network load which gives the PLM-ABB the best high sensitivity
range among the three measurement techniques here presented. Referring
to Fig. \ref{fig:ABB}, $Z_{PL}$ can be derived after simple computations
as 
\begin{equation}
Z_{PL}=-R_{sh}\frac{V_{a}}{V_{b}}.\label{eq:zpln_abb}
\end{equation}

All the aforementioned techniques can be designed to work in the PLC
frequency range (3 kHz -- 86 MHz). Since the two voltage measurements
needed can be performed by the same volt meter in different time instants,
the PLM receiver is enough to absolve this task. Considering the PLM-VNA
case, if the modem is used at the same time for full-duplex communication
\cite{7505646} and impedance measurement, the PLM receiver can measure
the network impedance only when no analog echo cancellation circuit
\cite{910475} is integrated in the modem. When analog cancellation
is present, it highly corrupts $V_{b}$ by rendering $\rho\simeq0$
in the whole frequency range. Hence, either the analog echo canceler
is fully characterized so that its transfer function can be digitally
removed, or a second receiver has to be integrated in the modem to
sense the signal before the analog echo canceler.

\subsection{Influence of the noise }

\begin{figure}[tb]
\begin{centering}
\subfloat[PLM-VNA\label{fig:VNA_n}]{\centering{}\includegraphics[width=0.7\columnwidth]{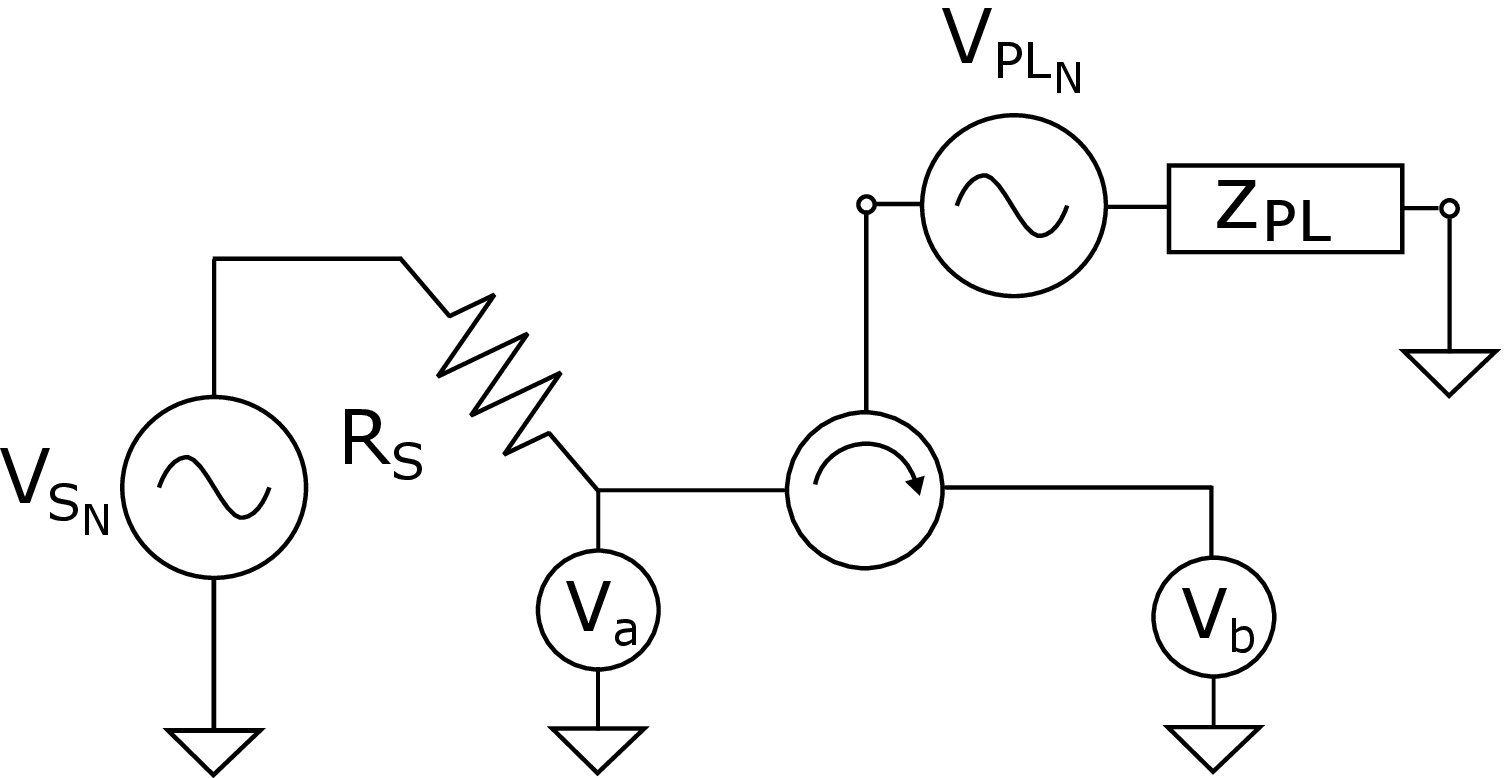}}\hfill{}\subfloat[PLM-IV\label{fig:RF-I-Vn}]{\centering{}\includegraphics[width=0.75\columnwidth]{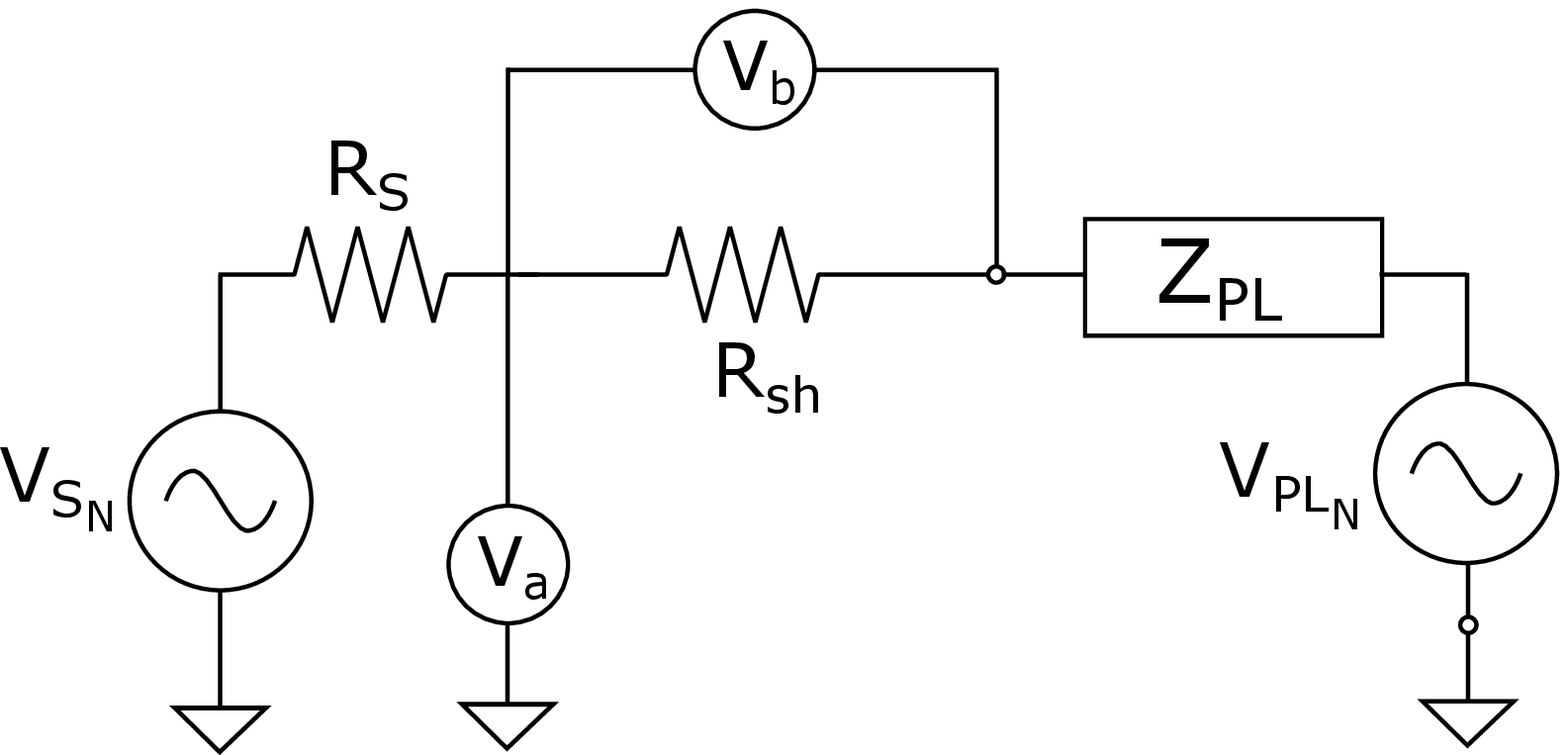}}\hfill{}\subfloat[PLM-ABB\label{fig:ABBn}]{\centering{}\includegraphics[width=0.8\columnwidth]{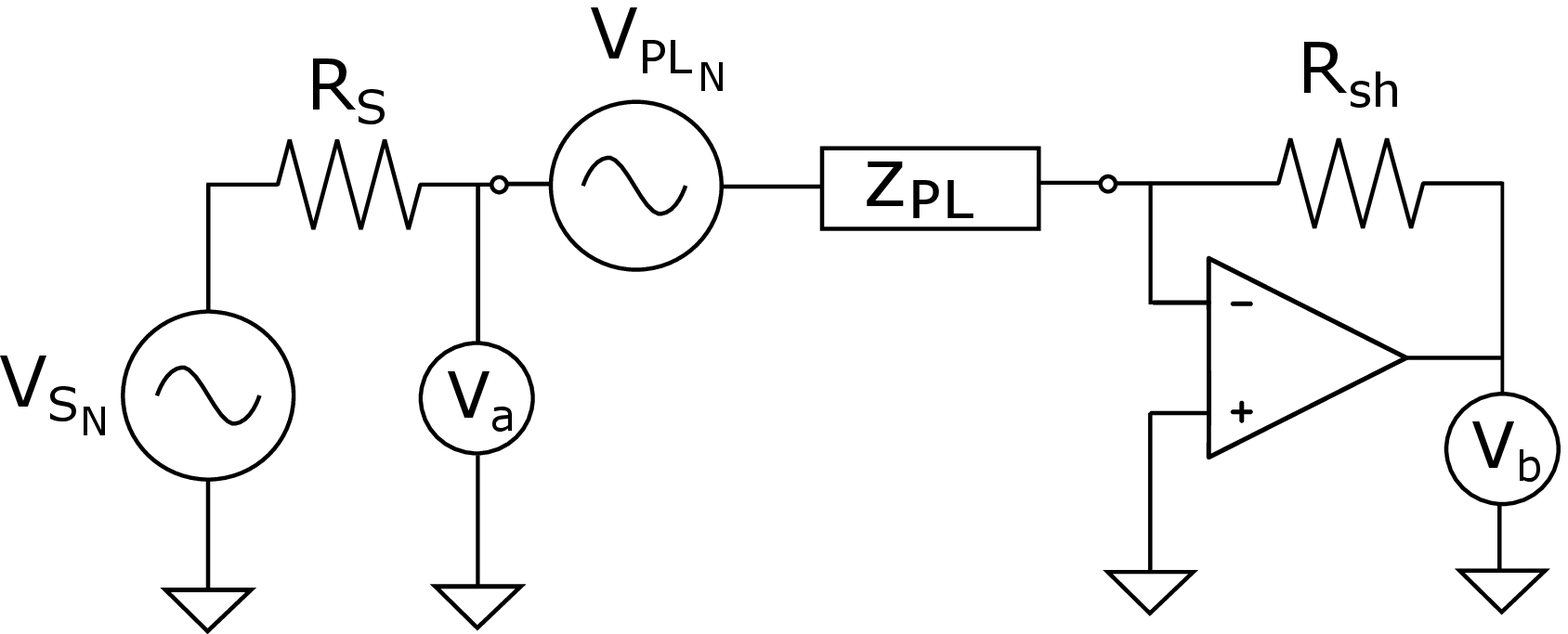}}
\par\end{centering}

\caption{Noise model of the selected impedance measurement methods.\label{fig:Noise-model-of}}
\end{figure}

In the following, we consider three sources of noise (see Fig. \ref{fig:Noise-model-of}):
$V_{S_{N}}$, generated by the transmitter as sum of the DAC, the
output filter and the line driver noises; $V_{R_{N}}$, generated
by the ADC at the receiver; $V_{PL{}_{N}}$, the equivalent background
noise coming from the network \cite{antonialirumore}. We decided
not to include in the following computations the equivalent noise
generated by $R_{sh}$, since it only accounts for the noise of one
resistor and is negligible compared to $V_{S_{N}}$, $V_{R_{N}}$
and $V_{PL{}_{N}}$. As already stated, the noise components due to
the non-ideality of the circulator and the opamp are not considered.

Considering as before an ideal circulator, the noisy measurement using
the PLM-VNA provides
\begin{subequations}
\begin{equation}
V_{a}=V_{a_{0}}+\frac{Z_{tr\_circ}}{Z_{tr\_circ}+R_{S}}V_{S_{N}}+V_{R_{Na}}\label{eq:vavna_n}
\end{equation}
\begin{equation}
V_{b}=V_{b_{0}}+\frac{Z_{tr\_circ}}{Z_{tr\_circ}+R_{S}}\rho V_{S_{N}}+\frac{Z_{o\_circ}}{Z_{o\_circ}+Z_{PL}}V_{PL{}_{N}}+V_{R_{Nb}}\label{eq:vbvna_n}
\end{equation}
\end{subequations}
where $V_{a_{0}}$ and $V_{b_{0}}$ are the measurements
in the absence of noise as described in Section \ref{sub:Measurement-equations},
and $Z_{tr\_circ}$ is the port impedance of the circulator at the
transmitter side. Using \eqref{eq:vavna_n}, \eqref{eq:vbvna_n} and
\eqref{eq:zpln_vna}, the measured value $\bar{Z}_{PL}$ of $Z_{PL}$
becomes
\begin{multline}
\bar{Z}_{PL}=Z_{o\_circ}\frac{V_{a_{0}}+V_{a_{0}}}{V_{a}-V_{b}}+\\
+\frac{2Z_{tr\_circ}Z_{PL}Z_{o\_circ}}{\left(V_{a}-V_{b}\right)\left(Z_{o\_circ}+Z_{PL}\right)\left(Z_{tr\_circ}+R_{S}\right)}V_{S_{N}}+\\
+\frac{Z_{o\_circ}^{2}}{\left(V_{a}-V_{b}\right)\left(Z_{o\_circ}+Z_{PL}\right)}V_{PL{}_{N}}+\frac{V_{R_{Na}}+V_{R_{Nb}}}{V_{a}-V_{b}}.\label{eq:zvna_n}
\end{multline}
Moreover, if $V_{a_{0}}\gg\left\{ V_{S_{N}},V_{PL{}_{N}},V_{R_{Na}},V_{R_{Nb}}\right\} $
and $Z_{PL}\lesssim Z_{o\_circ}$, then \eqref{eq:zvna_n} can be
approximated as 
\begin{multline}
\bar{Z}_{PL}\simeq Z_{PL{}_{0}}+\frac{Z_{tr\_circ}Z_{PL}}{V_{a_{0}}\left(Z_{tr\_circ}+R_{S}\right)}V_{S_{N}}+\\
+\frac{Z_{o\_circ}}{2V_{a_{0}}}V_{PL{}_{N}}+\frac{V_{R_{Na}}+V_{R_{Nb}}}{V_{a_{0}}(1-\rho)}=Z_{PL_{0}}+Z_{PL{}_{N}},\label{eq:zvna_n-1}
\end{multline}
where $Z_{PL{}_{0}}$ is the ideal network impedance computed using
\eqref{eq:zpln_vna}. The above formula shows that the transmitter
noise is linearly dependent on $Z_{PL}$ and inversely proportional
on $R_{S}$. The former result is due to the fact that a higher value
of $Z_{PL}$ causes a higher reflection of the transmitted noise to
the receiver; the latter result is due to the voltage partition principle.
The amplification of the network noise does not depend on the mismatch
between $Z_{PL}$ and $Z_{o\_circ}$, but only on $Z_{o\_circ}$.
That is because both the measured $V_{PL_{N}}$ and $V_{b_{0}}$ are
inversely proportional to $\left(Z_{o\_circ}+Z_{PL}\right)$, but
while $V_{PL_{N}}$ is inversely proportional to $V_{a_{0}}$, $V_{b_{0}}$
is directly proportional to it. Hence, the effect of $\left(Z_{o\_circ}+Z_{PL}\right)$
gets canceled.

As for the PLM-IV measurement technique (see Fig. \ref{fig:RF-I-Vn}),
the measured voltages are

\begin{subequations}
\begin{multline}
V_{a}=V_{a_{0}}+\frac{R_{sh}+Z_{PL}}{R_{sh}+Z_{PL}+R_{S}}V_{S_{N}}+\\
+\frac{R_{S}}{R_{sh}+Z_{PL}+R_{S}}V_{PL{}_{N}}+V_{R_{Na}}\label{eq:varfiv_n}
\end{multline}
\begin{multline}
V_{b}=V_{b_{0}}+\frac{R_{sh}}{R_{sh}+Z_{PL}+R_{S}}V_{S_{N}}-\\
-\frac{R_{sh}}{R_{sh}+Z_{PL}+R_{S}}V_{PL{}_{N}}+V_{R_{Nb}}.\label{eq:vbrfiv_n}
\end{multline}
\end{subequations} Using \eqref{eq:varfiv_n}, \eqref{eq:vbrfiv_n}
and \eqref{eq:zpln_rfiv}, $\bar{Z}_{PL}$ can be written as
\begin{multline}
\bar{Z}_{PL}=R_{sh}\left(\frac{V_{a_{0}}-V_{b_{0}}}{V_{b}}+\frac{Z_{PL}}{V_{b}\left(R_{sh}+Z_{PL}+R_{S}\right)}V_{S_{N}}\right)+\\
+R_{sh}\left(\frac{\left(R_{S}+R_{sh}\right)}{V_{b}\left(R_{sh}+Z_{PL}+R_{S}\right)}V_{PL_{N}}+\frac{V_{R_{Na}}-V_{R_{Nb}}}{V_{b}}\right).\label{eq:zrfiv_n}
\end{multline}
When the noise sources are much lower than the signal, \eqref{eq:zrfiv_n}
can be approximated as
\begin{multline}
\bar{Z}_{PL}\simeq Z_{PL{}_{0}}+\frac{Z_{PL}\left(R_{sh}+Z_{PL}\right)}{\left(R_{sh}+Z_{PL}+R_{S}\right)V_{a_{0}}}V_{S_{N}}+\\
+\frac{\left(R_{S}+R_{sh}\right)\left(R_{sh}+Z_{PL}\right)}{\left(R_{sh}+Z_{PL}+R_{S}\right)V_{a_{0}}}V_{PL{}_{N}}\\
+\frac{\left(R_{sh}+Z_{PL}\right)}{V_{a_{0}}}\left(V_{R_{Na}}-V_{R_{Nb}}\right)=Z_{PL{}_{0}}+Z_{PL{}_{N}}.\label{eq:zrfiv_n-1}
\end{multline}
The above formula shows that the transmitter noise, as for the PLM-VNA
case, increases almost linearly with $Z_{PL}$, while the network
noise has a Sigmoid dependence on its logarithm. Interestingly, the
transmitter noise is almost not influenced by $R_{sh}$ and it is
attenuated, especially for low values of $Z_{PL},$ as an inverse
function of $R_{S}$. On the other side, the channel noise increases
almost linearly with $R_{sh}$ and, for high values of $Z_{PL},$
also linearly with $R_{S}$. From these considerations, it emerges
that, in order to minimize the noise in the PLM-IV impedance measurement
technique, $R_{sh}$ has to be tuned to a small value, while a balance
has to be found for $R_{S}$ in order to minimize the total noise.

The noisy measurements performed with the PLM-ABB technique (see Fig.
\ref{fig:ABBn}) are

\begin{subequations}
\begin{equation}
V_{a}=V_{a_{0}}+\frac{Z_{PL}V_{S_{N}}+R_{S}V_{PL{}_{N}}}{Z_{PL}+R_{S}}+V_{R_{Na}}\label{eq:vaabb_n}
\end{equation}
\begin{equation}
V_{b}=V_{b_{\text{0}}}-\frac{R_{sh}\left(V_{S_{N}}+V_{PL{}_{N}}\right)}{Z_{PL}+R_{S}}+V_{R_{Nb}},\label{eq:vbabb_n}
\end{equation}
\end{subequations}which, combined with \eqref{eq:zpln_abb}, give
\begin{multline}
\bar{Z}_{PL}=R_{sh}\left(\frac{V_{a_{0}}}{\left(Z_{PL}+R_{S}\right)V_{b}}+\frac{Z_{PL}}{\left(Z_{PL}+R_{S}\right)V_{b}}V_{S_{N}}\right)+\\
+R_{sh}\left(\frac{R_{S}}{\left(Z_{PL}+R_{S}\right)V_{b}}V_{PL{}_{N}}+\frac{V_{R_{Na}}}{V_{b}}\right).\label{eq:zabb_n}
\end{multline}
When the noise sources are much lower than $V_{a}$, \eqref{eq:zabb_n}
can be approximated as
\begin{equation}
\bar{Z}_{PL}=Z_{PL_{0}}-\frac{Z_{PL}^{2}}{V_{a_{0}}}V_{S_{N}}-\frac{R_{S}Z_{PL}}{V_{a_{0}}}V_{PL{}_{N}}+\frac{Z_{PL}}{V_{a_{0}}}V_{R_{Na}}.\label{eq:zabb_n-1}
\end{equation}
This formula shows that the noise is indipendent of $R_{sh}$. This
is due to the decoupling action of the I-V converter. On the other
side, the transmitted noise has a quadratic dependence on $Z_{PL}$
and the network noise is linearly dependent both on $Z_{PL}$ and
$R_{S}$. 

The dependence of the impedance measurement noise on the circuit parameters
are finally summarized in Table \ref{tab:Qualitative-dependence-of}.
\begin{table}[tb]
\begin{centering}
\caption{Qualitative dependence of ${Z}_{PL_N}$ on the circuit parameters.\label{tab:Qualitative-dependence-of} }

\par\end{centering}

\centering{}%
\begin{tabular}{|c|c|c|c|c|}
\hline 
 & $V_{a_{0}}$ & $Z_{PL}$ & $R_{S}$ & $R_{sh}$ or $Z_{o\_circ}$\tabularnewline
\hline 
\hline 
VNA & inverse & linear--indep. & inverse--indep. & indep.--linear\tabularnewline
\hline 
RF I-V & inverse  & linear--sigmoid & inverse--linear & indep.--linear\tabularnewline
\hline 
ABB & inverse  & quadrat.--linear & indep.--linear & indep.\tabularnewline
\hline 
\end{tabular}
\end{table}

\subsection{Modeling of the ADC and DAC noise\label{sub:Modeling-of-the}}

As presented in the previous subsections, the measure of $Z_{PL}$
 involves the transformation of a digital test signal to analog domain
and the computation of the ratio of two measured quantities. This
operation can be performed by an analog divider circuit, which however
is in most cases very expensive and still provides an error in the
order of 1\%. Hence, we propose to perform the division in digital
domain, whose precision depends almost exclusively on the ADC resolution.
To this aim, we already included in the previous calculations both
the transmitter output noise $V_{S_{N}}$ and the acquisition
 noise $V_{R_{N}}$ for $V_{a}$ and $V_{b}$.
The influence of these noise sources on the measurements is different
based on the measurement approach considered. We differentiate between
two different measurement approaches: sequential and cumulated. 

In the first approach, pure tones are transmitted sequentially, and
a different measurement is performed for every single tone. In this
case, the DAC and the ADC can be tuned to exactly clip at the peak amplitude of the transmitted
and received tone respectively. The resulting signal-to-quantization-noise
ratio (SQNR) is simply computed from the effective number of bits
of the DAC or the ADC. Taking into account also distortion effects
of the whole TX and RX chains, PLMs equipped with a 12-bit DAC and a 12-bit ADC, can reach a signal-to-noise-and-distortion
ratio (SINAD) of 69 dB at the end of both the TX and RX chains\cite{BB_frontend}.

In the second approach, the test signal is transmitted over all the
test frequencies at the same time, and a single broad-band measurement
is performed. This is the approach of the orthogonal frequency division
multiplexing (OFDM) systems typically deployed in PLC. Although the
average generated power may be the same over all the OFDM tones,
the peak power of the OFDM signal is much higher than the average power, thus leading to an high peak-to-average-power
ratio (PAPR). Hence, the balancing between clipping and quantization
noise highly reduces the SINAD, so that it reaches 60 dB for a 12-bit
ADC with an optimum tuning of the parameters \cite{4429546,Berger:11}.

The sequential approach yields the best SINAD but it comes at cost
of a much slower acquisition time, whereas the second exploits the
features of the PLM to perform a single measurement at a cost of worse
SINAD. We remark that in both approaches the transmission time has
to be set in order to let the whole network resonate to the transmitted
tones. In the following simulations, we focus on the best achievable
measurement performance and hence we rely on the sequential approach.
The possible use of the cumulated approach is also briefly discussed.

\section{Reflection coefficient and FDR measurements \label{sec:FDR-measurement-technique}}

In this section, we briefly analyze a technique to measure $\rho$
and $T$ in order to allow a fair comparison with the impedance measurement
techniques presented in Section \ref{sec:Impedance-measurement-techniques}.

A direct way to measure $\rho$ and $T$ is to rely on the VNA circuit
architecture depicted in Fig. \ref{fig:Reflectometer-(VNA-or}, where
the generated signal is forwarded to the network through a hybrid
coupler.  Different techniques can be applied to properly form
the transmitted signal, including the two approaches presented in
Section \ref{sub:Modeling-of-the}, and to analyze the received one
\cite{thesisMont}. In our approach, the signals $V_{a}$ and $V_{b}$
are firstly digitalized and then \eqref{eq:rho-1} and \eqref{eq:trace}
are applied to derive $\rho$ and $T$ respectively. 

When noise is present, as depicted in Fig. \ref{fig:VNA_n}, \eqref{eq:vavna_n}
and \eqref{eq:vbvna_n} apply. Hence, the resulting noisy reflection
coefficient and trace are
\begin{multline}
\bar{\rho}=\frac{V_{b}}{V_{a}}\simeq\rho_{0}+\frac{\rho Z_{tr}}{\left(Z_{tr}+R_{S}\right)V_{a_{0}}}V_{S_{N}}+V_{R_{Nb}}+\\
+\frac{Z_{o\_circ}}{\left(Z_{o\_circ}+Z_{PL}\right)V_{a_{0}}}V_{PL{}_{N}}=\rho_{0}+\rho_{N}\label{eq:rho_noise}
\end{multline}
 
\begin{multline}
\bar{T}=V_{a}V_{b}\simeq\rho\left|V_{a_{0}}\right|^{2}+2\rho\frac{Z_{tr\_circ}}{Z_{tr\_circ}+R_{S}}V_{a_{0}}V_{S_{N}}+\\
\rho V_{a_{0}}V_{R_{Na}}+\frac{Z_{o\_circ}}{Z_{o\_circ}+Z_{PL}}V_{a_{0}}V_{PL{}_{N}}+V_{a_{0}}V_{R_{Nb}}=T_{0}+T_{N}\label{eq:T_noise}
\end{multline}
respectively, taking into account the same approximation used for
\eqref{eq:zvna_n-1}. Here $\rho_{0}$ and $T_{0}$ are the quantities
of interest, while $\rho_{N}$ and $T_{N}$ are the noise components. 

Also in this case the noise is filtered by the presence of the circulator
characteristic impedance, the transmitter output impedance and the
channel impedance. Conversely from impedance measurements, in the
$\bar{T}$ measurement case $V_{a_{0}}$ is now a multiplicative factor
for the noise.  We remark that, although the error propagation in
\eqref{eq:rho_noise} and \eqref{eq:T_noise} is similar, the noise
component related to $V_{S_{N}}$ has double amplitude in the $\bar{T}$
measurement. Moreover, $T_{0}$ is function of the actual reflection
coefficient $\rho$, while $\rho_{0}$ is only approximately equal
to $\rho$. The difference is due to the noise caused by $V_{S_{N}}$
and $V_{R_{Na}}$ in the voltage division and is not made explicit
in \eqref{eq:rho_noise}, since it is negligible in most of the cases.
The effect of these aspects on the overall noise is investigated in
detail in Section \ref{sec:Comparison-of-the}.

\section{Considerations on the PLC features\label{sec:Considerations-on-the}}

In order to assess the performance of the measurement techniques presented
in this paper, a closer insight on the PLC characteristics is needed,
since PLM operate on frequency ranges and with transmitted power that
are specified by PLC standards. 

The PL medium has some particular features that differentiate it from
wired transmission links like telephone twisted-pair cables. First
of all, it is made of power cables whose physical characteristics
have not been standardized; this means that any cable in a network
may have different electrical characteristics that cannot be classified
(unlike AWG cables). Secondly, the power cords are not twisted and,
more in general, not protected versus possible electro-magnetic interference.
Lastly, the loads branched to the PLNs are not matched at PLC frequencies;
moreover, they are often time-varying and also produce both thermal
and impulsive noise that can be either periodic or not \cite{5764408}.
For theses reasons, the front-end of PLC modems is generally engineered
to maximize the voltage sent to the network or received from it \cite{7407357},
instead of maximizing the sent/received power, as in DSL \cite{DSL}.
Hence, the output impedance $R_{S}$ of both narrowband and broadband
PLC modems has commonly a value ranging from hundreds of m$\Omega$s
to few $\Omega$s.

Since $R_{S}$ has a very low value, the transmitted signal has the
same order of magnitude of the generated signal. In order to save
power at the transmitter and to send the maximum voltage possible
to the network, also the value of $R_{sh}$ has to be lower or at
maximum comparable to $Z_{PL}$. In the case of the PLM-VNA, it has
been shown that to optimize the communication throughput, $Z_{o\_circ}$
needs to be matched to the average $Z_{PL}$ \cite{7476291}. 

As already said, $Z_{PL}$ is both frequency and time varying. While
in most of the scenarios it has a periodic time variation that follows
the mains cycle, the frequency variation is due to the topological
characteristics of the network and can have very complex shapes. In
in-home environments, the absolute value of the broadband (2-86 MHz)
impedance normally ranges from 50 $\Omega$ to 300 $\Omega$ \cite{6620966},
while in low-voltage distribution networks the absolute value of the
narrowband (3-500 kHz) impedance can go down to few $\Omega$ \cite{s8128027}.

Finally, $V_{PL_{N}}$ is a major issue in PLCs because of its impulsive
nature and possibly high value. However, the impulsive noise component of $V_{PL_{N}}$
might not necessarily compromise the accuracy of the measurement. 
In fact, impedance measurements can be performed
during the time windows where impulsive noise is not present and its
transients are also ended. Otherwise, their effect can be mitigated
by averaging many measurements over time. As for the background noise,
it can be described as a complex colored Gaussian process (CCGN),
with power spectrum inversely proportional to the frequency. Different
measurement campaigns have been performed to measure the noise both
in MV distribution lines \cite{4265730}, LV distribution lines \cite{4039493}
and in-home environments \cite{5479897}. They show that the background
noise spans on average from -60 dBm/Hz to -110 dBm/Hz in the narrowband,
and from -110 dBm/Hz to -140 dBm/Hz in the broadband. This information
is useful when compared to the transmitted power allowed by the communications
standards. 

The IEEE standard for narrowband PLC limits the maximum voltage injected
into the network in the Cenelec frequencies (3 to 148.5 kHz) in the
range 97 dB$\mu$V/Hz to 110 dB$\mu$V/Hz, based on the frequency
sub-band \cite{nbstd}. The same standard also limits the maximum
power that can be transmitted in the FCC band (150 kHz to 500 kHz)
in the range -45 dBm/Hz to -55 dBm/Hz, depending on the frequency.
In reference to broadband PLC, the IEEE standard \cite{bbstd} limits the
maximum power that can be transmitted to -55 dBm/Hz in the 2 MHz -
30 MHz band and to -85 dBm/Hz in the 30 MHz - 86 MHz band. All the
aforementioned values are referred to a standard load impedance whose
value is $R_{SL}=$50 $\Omega$ for the FCC bands and the broadband
PLC, and $Z_{SL}=$50 $\Omega$ \textbackslash{}\textbackslash{} (5
$\Omega$ + 50 $\mu$H) for the Cenelec bands \cite{cenelec}, where
\textbackslash{}\textbackslash{} denotes the parallel operator on
two impedances.

\section{Comparison of the measurement techniques\label{sec:Comparison-of-the}}

Considering all the relations presented in Sections \ref{sec:Impedance-measurement-techniques}
and \ref{sec:FDR-measurement-technique}, we can now compare the three
impedance measurement techniques, the $\rho$ measurement and the
FDR techniques in terms of overall error. In particular, we define
the quantity (of interest)-to-noise ratio (QNR) as 
\begin{equation}
QNR=\frac{E\left[\left|X_{0}\right|^{2}\right]}{E\left[\left|X_{N}\right|^{2}\right]},\label{eq:inr}
\end{equation}
where $E\left[\cdotp\right]$ is the expectation operator and $X$
can be either $Z_{PL}$, $\rho$, or $T$. This relation is more useful
than the simple noise variance, since it normalizes the noise variance
with the magnitude of the true data, thus providing an easily comparable
result. 

In order to set a fair comparison, we consider the transmitters of
the three circuits under analysis to produce the same voltage and
power on the test load $R_{SL}$. This causes the three methods to
generate different $V_{S}$, since the output impedance towards the
network is different. In particular, using the same value of $R_{S}$,
the ABB method generates the lowest $V_{S}$. It is then the one that
generates the lowest $V_{S_{N}}$ and consumes less power. 

To have a uniform voltage and power reference for the whole PLC bands,
we use $R_{SL}$ as a common impedance reference. Since the Cenelec
bands use the reference load $Z_{SL}$ instead (see Section \ref{sec:Considerations-on-the}),
their maximum allowed voltage has to be converted. The limit voltage
$V_{max\text{\_R}}$ for the Cenelec bands referred to $R_{SL}$ is
\begin{equation}
V_{max\text{\_R}}=\frac{R_{SL}}{R_{SL}+R_{Scmp}}\frac{Z_{SL}+R_{Scmp}}{Z_{SL}}V_{max\_Z},
\end{equation}
where $V_{max\_Z}$ is the limit voltage for the Cenelec bands referred
to $Z_{SL}$ and $R_{Scmp}$ is the overall output impedance of the
modem seen from the network. With this conversion, $V_{max\text{\_R}}$
ranges from 97 dB$\mu$V/Hz to 114 dB$\mu$V/Hz, or equivalently the
maximum transmitted power ranges from -10 dBm/Hz to 7 dBm/Hz.

As for the network noise, we generate it as shown in the Appendix.
The power spectral density needed for the noise generation can be
obtained from the power profiles presented in Section \ref{sec:Considerations-on-the}
as follows. The noise power profiles are generally referred to the
input impedance of a spectrum analyzer, which is $R_{SA}=$50$\Omega$.
If we call $P_{SA}$ the noise power measured by the spectrum analyzer
and $P_{PL_{N}}$ the average power of the network noise, then
\begin{equation}
P_{PL_{N}}=\frac{(R_{SA}+Z_{PL})}{R_{SA}}P_{SA}.
\end{equation}

As for $V_{S_{N}}$ and $V_{R_{N}}$, we assume the modem to be equipped
with a 10-bit DAC and a 12-bit ADC, so that the noise at the output
of the transmitter is 55 dB below $V_{S}$ \cite{NB_frontend} and
the noise at the end of the receiver chain is 69 dB below $V_{a}$ and
$V_{b}$ respectively.

In the following, \eqref{eq:zvna_n}, \eqref{eq:zrfiv_n}, \eqref{eq:zabb_n},
\eqref{eq:rho_noise} and \eqref{eq:T_noise} will be thoroughly analyzed
in different conditions.

\subsection{Comparison in terms of transmitted power}

In this subsection, we consider fixed impedance values and show how
the QNR varies as function of the channel noise power $P_{PL_{N}}$
and of the power $P_{TX}$ on $Z_{PL}$. In particular, we set $Z_{PL}=Z_{tr\_circ}=$
50 $\Omega$, $Z_{o\_circ}=$ 100 $\Omega$, $R_{sh}=R_{S}=$ 1 $\Omega$.

\begin{figure}[tb]
\subfloat[$P_{TX}=$-10 dBm]{\centering{}\includegraphics[width=0.5\columnwidth]{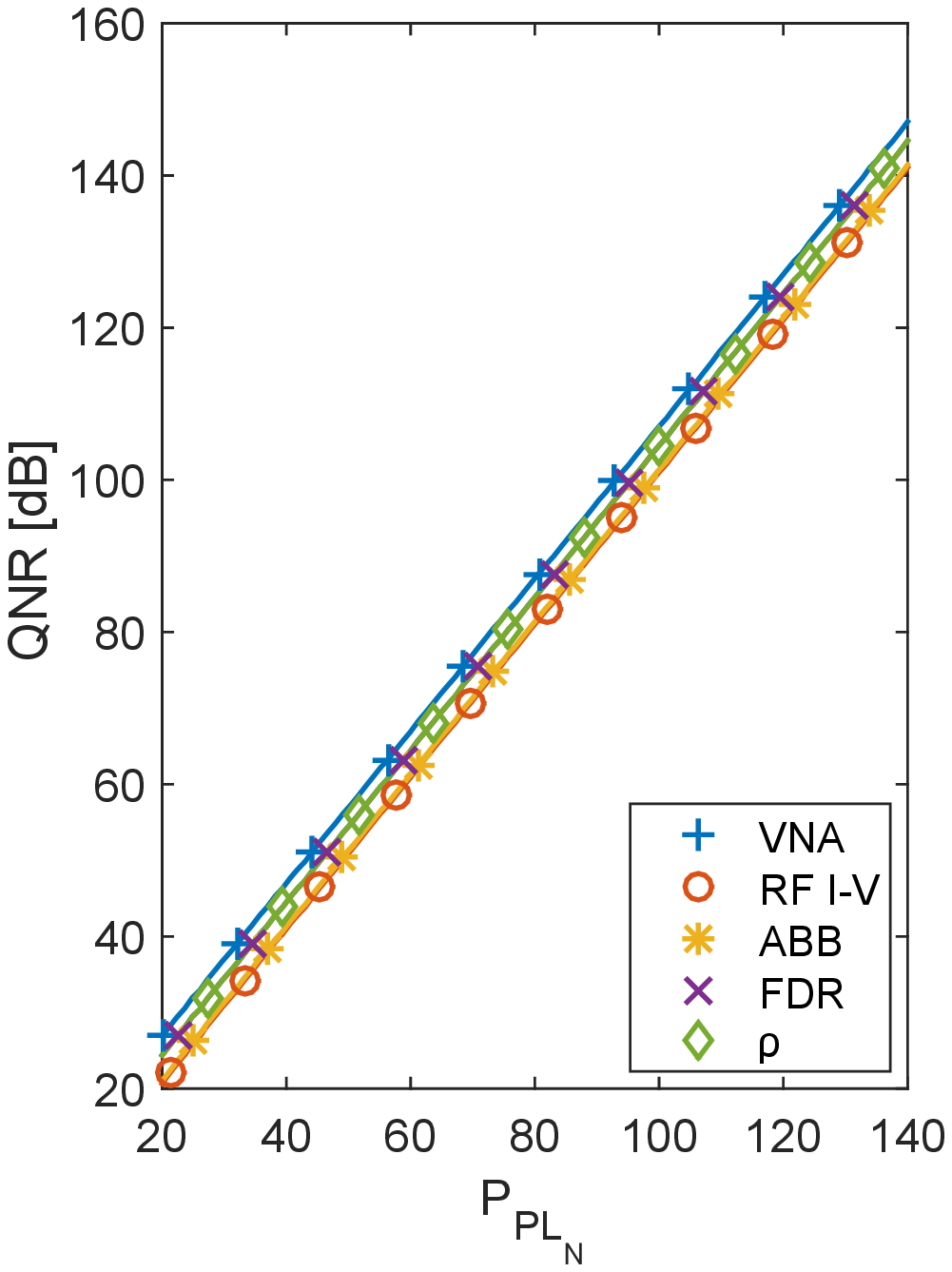}}\subfloat[$P_{PL_{N}}=$-60 dBm]{\centering{}\includegraphics[width=0.5\columnwidth]{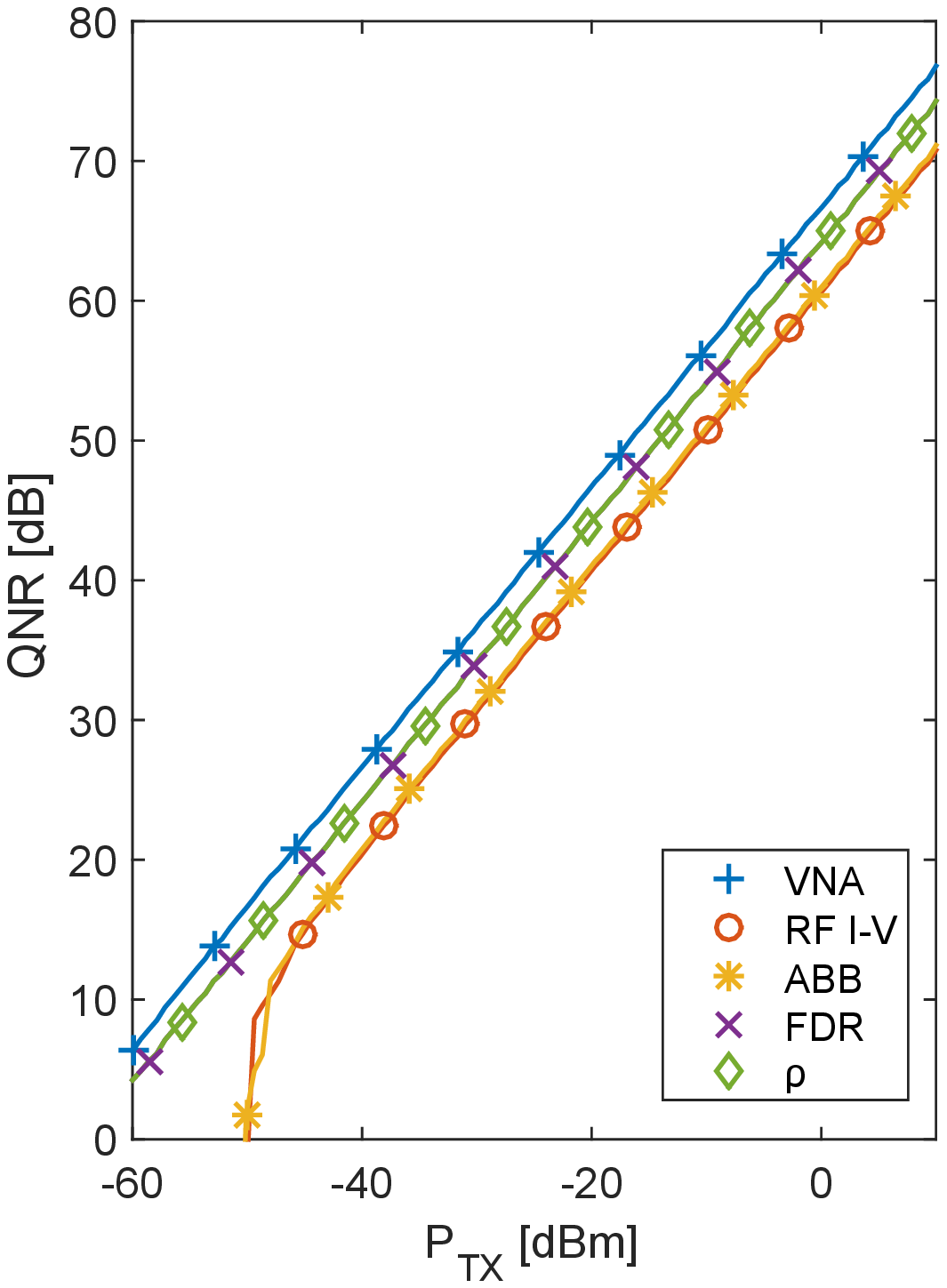}}\caption{QNRas function of $P_{TX}$ and $P_{PL_{N}}$, when $V_{S_{N}}=V_{R_{N}}=0$.\label{fig:INR-and-TNRideal}}
\end{figure}
\begin{figure}[tb]
\subfloat[$P_{TX}=$-10 dBm]{\centering{}\includegraphics[width=0.5\columnwidth]{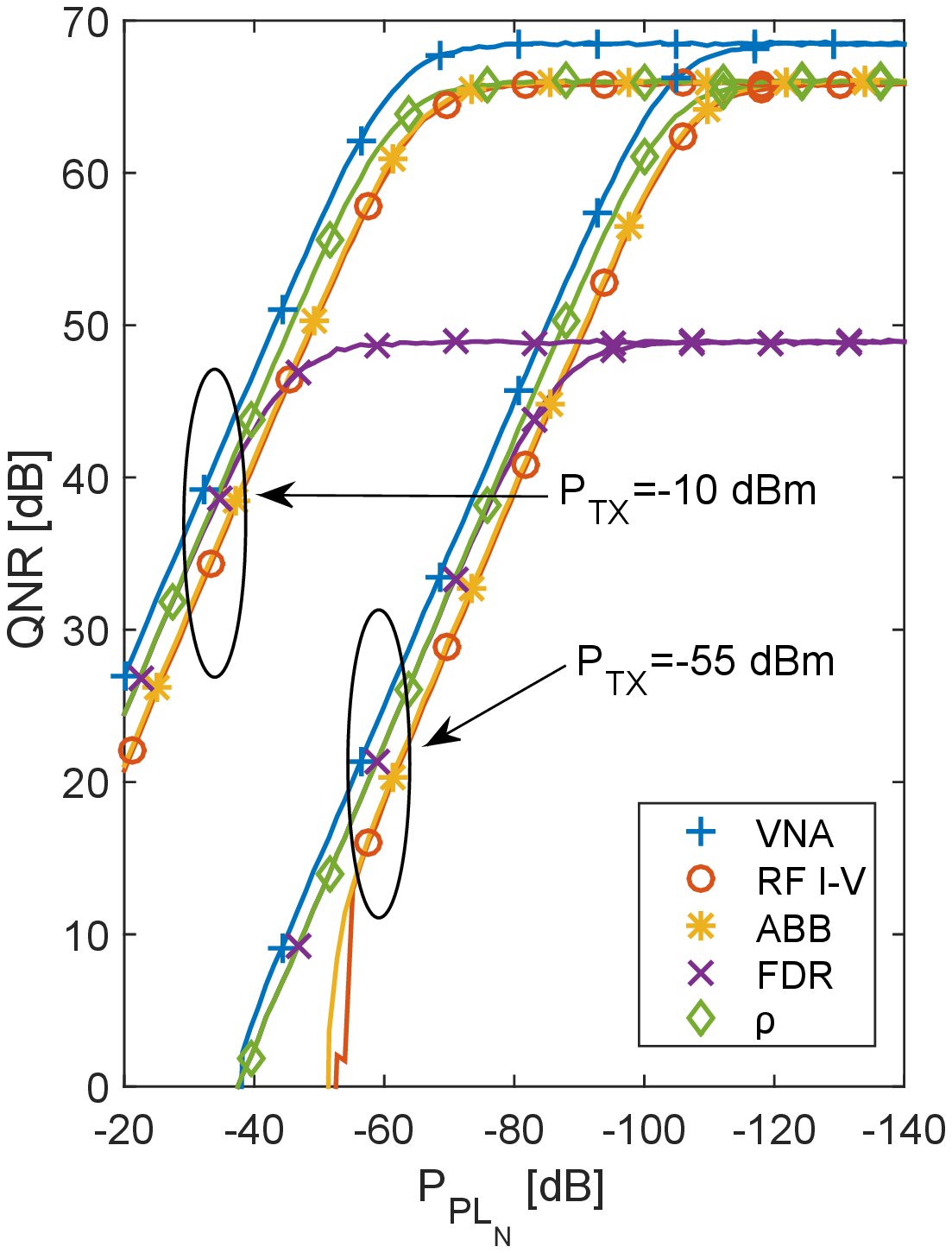}}\subfloat[$P_{PL_{N}}=$-70 dBm]{\centering{}\includegraphics[width=0.5\columnwidth]{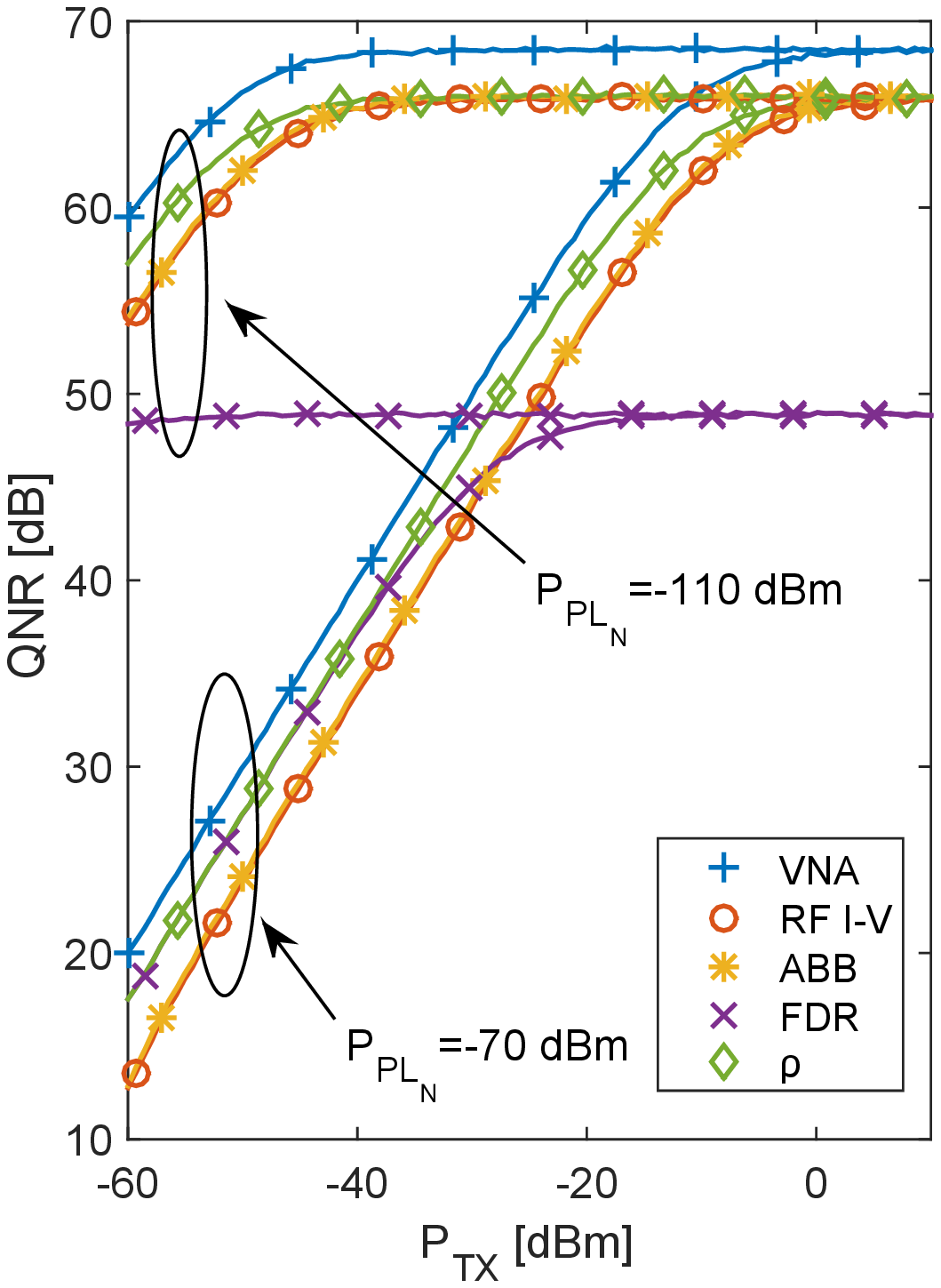}}\caption{QNR as function of $P_{TX}$ and $P_{PL_{N}}$, whith standard values
of $V_{S_{N}}$ and $V_{R_{N}}$.\label{fig:INR-and-TNR}}
\end{figure}
Fig. \ref{fig:INR-and-TNRideal} shows that, when no modem noise is
considered, the QNR has a linear trend both versus $P_{PL_{N}}$ and
$P_{TX}$, for all the considered methods. Moreover, the best QNR
is obtained with the PLM-VNA method. For high values of background
noise and low transmitted power though, the QNR of the PLM-ABB and
PLM-IV methods declines more rapidly. 

Considering now the ADC and DAC noise components (see Section \ref{sec:Considerations-on-the}), the situation
changes. The
plots of Fig. \ref{fig:INR-and-TNR} show the QNR trends for $P_{TX}$
and $P_{PL_{N}}$ values that are typical of narrowband ($P_{TX}=$-10
dBm, $P_{PL_{N}}=$-70 dBm) and broadband ($P_{TX}=$-55 dBm, $P_{PL_{N}}=$-110
dBm) PLC. All the techniques tend to saturate to the QNR for which
the contribution of $V_{S_{N}}$ becomes greater than that of $V_{PL_{N}}$.
The FDR curves saturate far below the others. This is due to the fact
that, according to \eqref{eq:T_noise}, $T_{0}$ is the exact value
of the FDR trace, while $\rho_{0}$ and $Z_{PL_{0}}$ measured with
all the considered techniques are actually perturbed by a small error
caused mostly by $V_{S_{N}}$ which is not rendered explicit in the
simplified equations \eqref{eq:zvna_n-1}, \eqref{eq:zrfiv_n-1},
\eqref{eq:zabb_n-1} and \eqref{eq:rho_noise}. This error is negligible
for low QNRs, but at higher QNRs tends to balance the noise contribution
due to $V_{S_{N}}$ in $\rho_{N}$ and $Z_{PL_{N}}$. Since $T_{0}$
is an exact value, the same balancing does not apply to $T_{N}$.

We finally point out that, using the cumulated measurement approach
instead of the sequential one, the ADC noise is increased, thus fixing
the saturation of the QNR to lower values.

\subsection{Comparison in terms of impedance value}

In this subsection, we consider fixed $P_{TX}$ and $P_{PL_{N}}$ values
and show how the QNR varies as function of $Z_{PL}$. In particular,
we set $P_{TX}=$ -10 dBm, $P_{PL_{N}}=$ -70 dBm, $R_{sh}=R_{S}=$
1 $\Omega$, $Z_{tr\_circ}=$ 50 $\Omega$, $Z_{o\_circ}=$ 100 $\Omega$.

\begin{figure}[tb]
\subfloat{\centering{}\includegraphics[width=0.5\columnwidth]{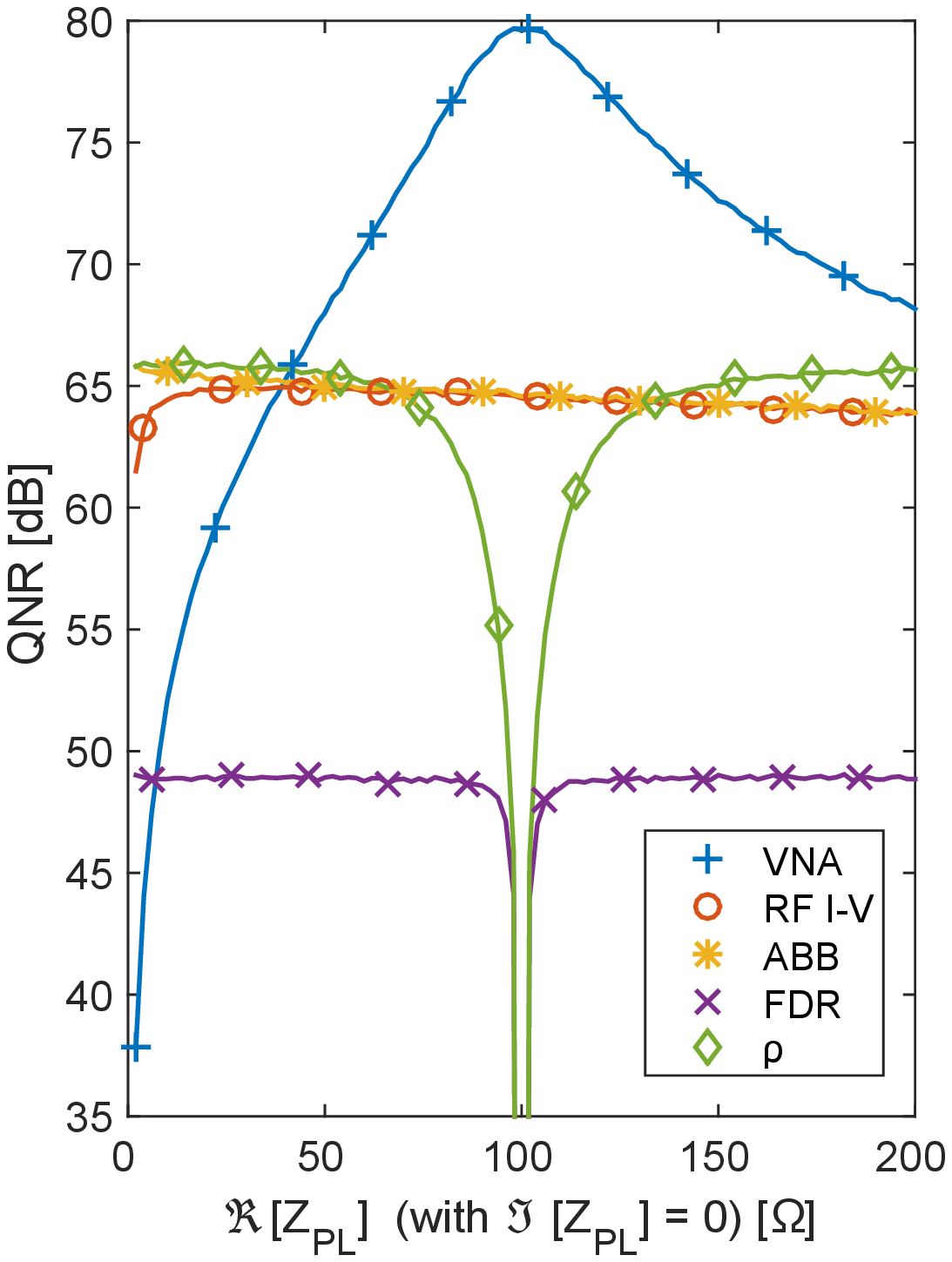}}\subfloat{\centering{}\includegraphics[width=0.5\columnwidth]{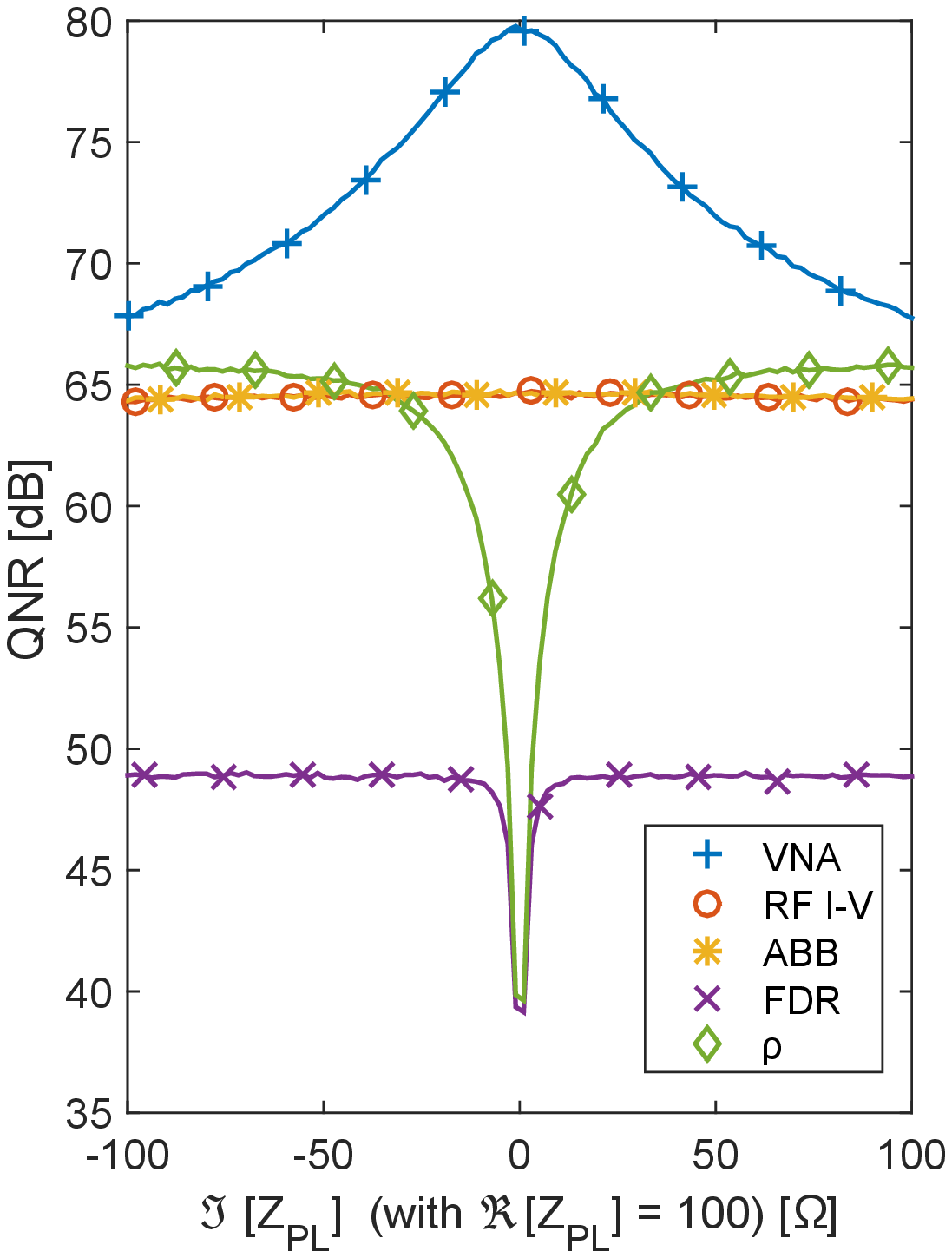}}\caption{QNR as function
of $Z_{PL}$, with $P_{TX}=-10$ dBm and $P_{PL_{N}}=-70$ dBm. \label{fig:INR-and-TNRImp}}
\end{figure}
According to the results of the previous subsection, for high impedance
values the PLM-VNA method has the highest QNR. However, Fig. \ref{fig:INR-and-TNRImp}
shows that its accuracy decreases for lower impedance values. This
is due to the fact that when $Z_{PL}\ll Z_{o\_circ}$ the numerator
in \eqref{eq:zpln_vna} becomes very small, thus enhancing the effect
of the noise. Conversely, the QNR of the PLM-ABB and PLM-IV methods
is maximum for low values and slowly decreases for higher values.
Below 10 $\Omega$ the PLM-ABB method has the best QNR, while the
QNR of the PLM-IV method shows a decreasing trend. This is due
to the fact that for this comparison we considered the configuration
that is more suited for high impedance measurements. Fig. \ref{fig:INR-and-TNRImp}
also shows that the QNR value in only slighted affected by the immaginary
part of $Z_{PL}$ for all the methods considered. As for the FDR and
$\rho$ measurement techniques, we notice that when $Z_{PL}=Z_{o\_circ}$
the QNR in dB goes to $-\infty$. This is due to the fact that in this case
$V_{b_{0}}=0$: since a multiplication or a division is performed
to derive T and $\rho$ respectively, this will result in a pure noise
component.

\subsection{Additive Gaussian noise limits}

\begin{figure}[tb]
\centering{}\subfloat[$P_{TX}=$-10 dBm]{\centering{}\includegraphics[width=0.5\columnwidth]{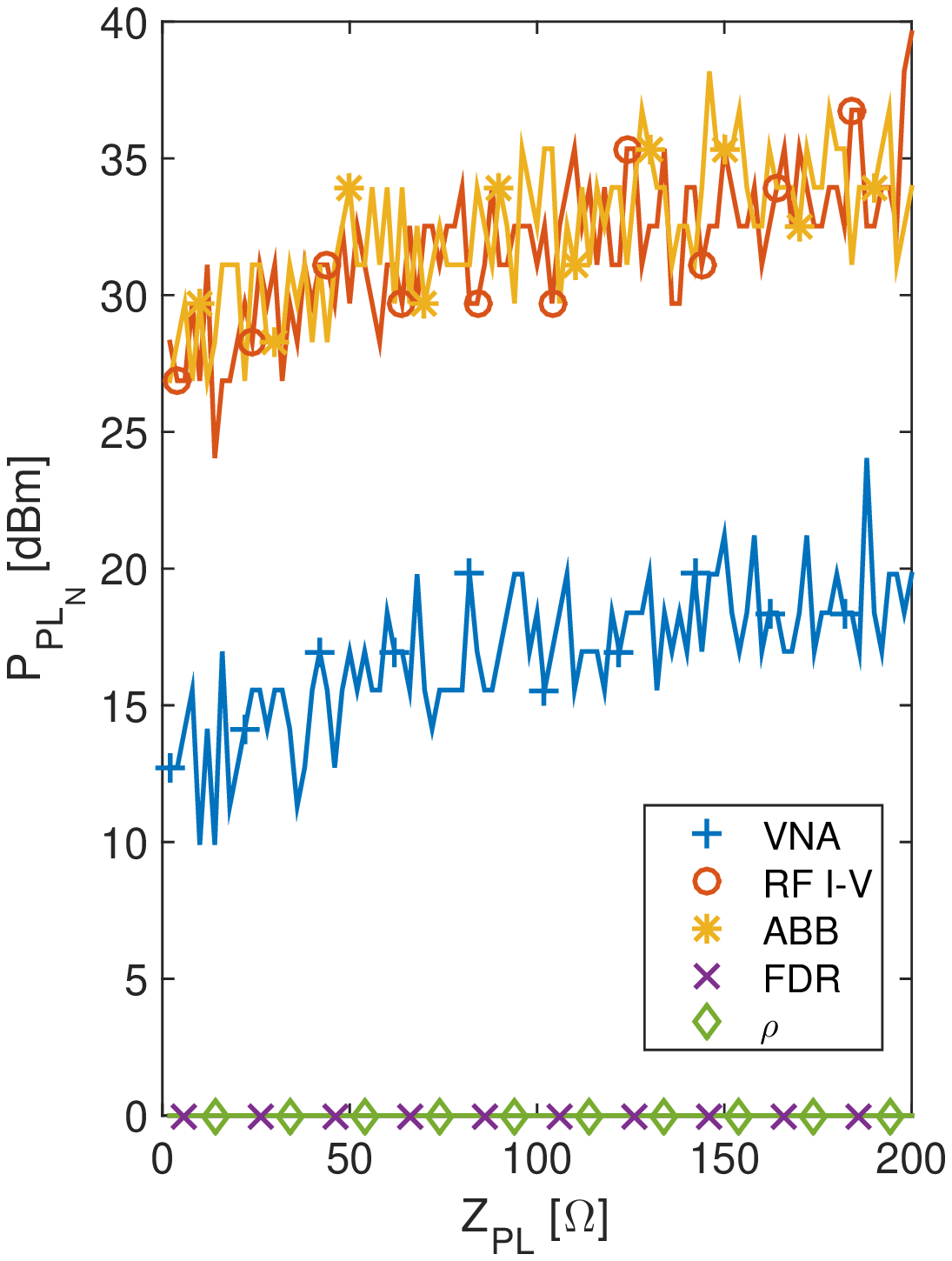}}\subfloat[$Z_{PL}=$100 $\Omega$]{\centering{}\includegraphics[width=0.5\columnwidth]{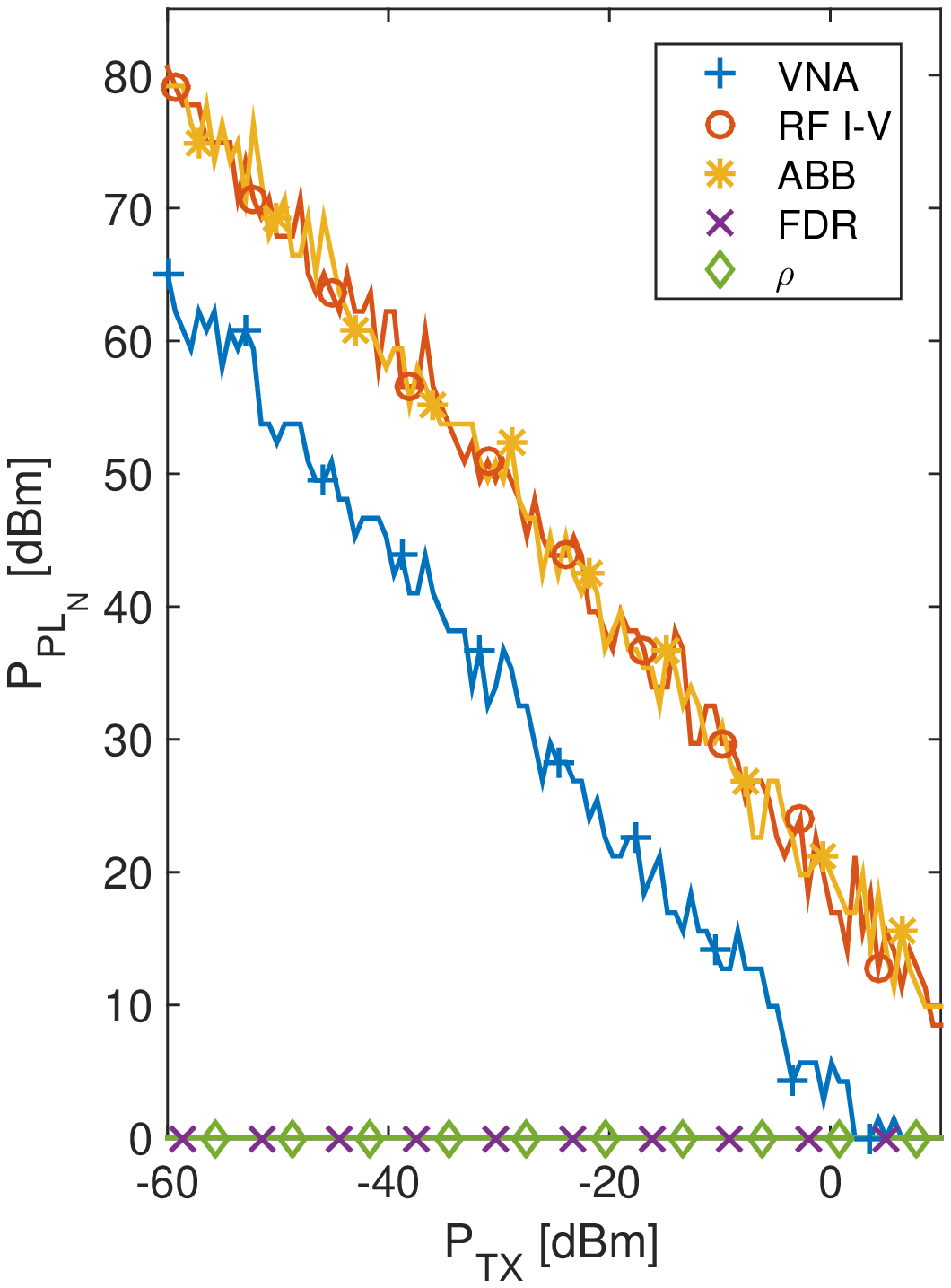}}\caption{Maximum $P_{PL_{N}}$ for which $Z_{PL_{N}}$ is Gaussian as function
of $Z_{PL_{0}}$ and $P_{TX}$.\label{fig:Maximum-for-which}}
\end{figure}

In this subsection, we analyze the noise limits for which the overall
noise of the considered measurement techniques can be considered additive
Gaussian, thus allowing the simplifications made in \eqref{eq:zvna_n-1},
\eqref{eq:zrfiv_n-1} and \eqref{eq:zabb_n-1}. To do so, we firstly
analyzed the mean of \eqref{eq:zvna_n}, \eqref{eq:zrfiv_n} and \eqref{eq:zabb_n}
using different parameter values and found that the noise is zero
mean for all the examined techniques when $P_{PL_{N}}[dBm]<P_{TX}[dBm]-10[dBm]$.
We also performed the classical Kolmogorov-Smirnov test \cite{daniel1990applied}
over the same equations to check their gaussianity. Fig. \ref{fig:Maximum-for-which}
shows the maximum power of the PL noise for which $\bar{Z}_{PL_{N}}$
is additive Gaussian. The maximum $P_{PL_{N}}$ is steeply proportional
to $P_{TX}$ and increases almost exponentially with $Z_{PL}$. However,
the limits shown are higher than the normal power line background
noise values. For example, if $P_{TX}=$ -10 dBm and $Z_{PL}=$100,
$\bar{Z}_{PL_{N}}$ measured with the PLM-ABB technique is perturbed
by additive Gaussian noise when $P_{PL_{N}}<$ -40 dBm, which is on
average true. We conclude that average impedance measurements performed
on power grids should provide results perturbed with additive Gaussian
noise.

The null value of the FDR method is again due to the fact that $T$
is the result of a multiplication \eqref{eq:T_noise}, yielding overall a mixed
sum of Gaussian and Chi-square distributions. When $P_{PL_{N}}$ is low, the DAC noise dominates
in the Chi-square pairs, which therefore degenerate to Gaussian variables.
The opposite is valid when $P_{PL_{N}}$ is high. Hence, although
the FDR technique provides the worst accuracy in terms of QNR, it
provides the best Gaussian profile, i.e. it is the simplest to analyze
and simplify.

\subsection{Application to fault detection}

In this subsection, we show the results obtained by applying all the
aforementioned measurement techniques to the problem of detecting
a fault in a Smart Grid. This way, we assess the performance of each
technique in a concrete application.

\begin{figure}[tb]
\centering{}\subfloat[Impedance based method]{\centering{}\includegraphics[width=0.5\columnwidth]{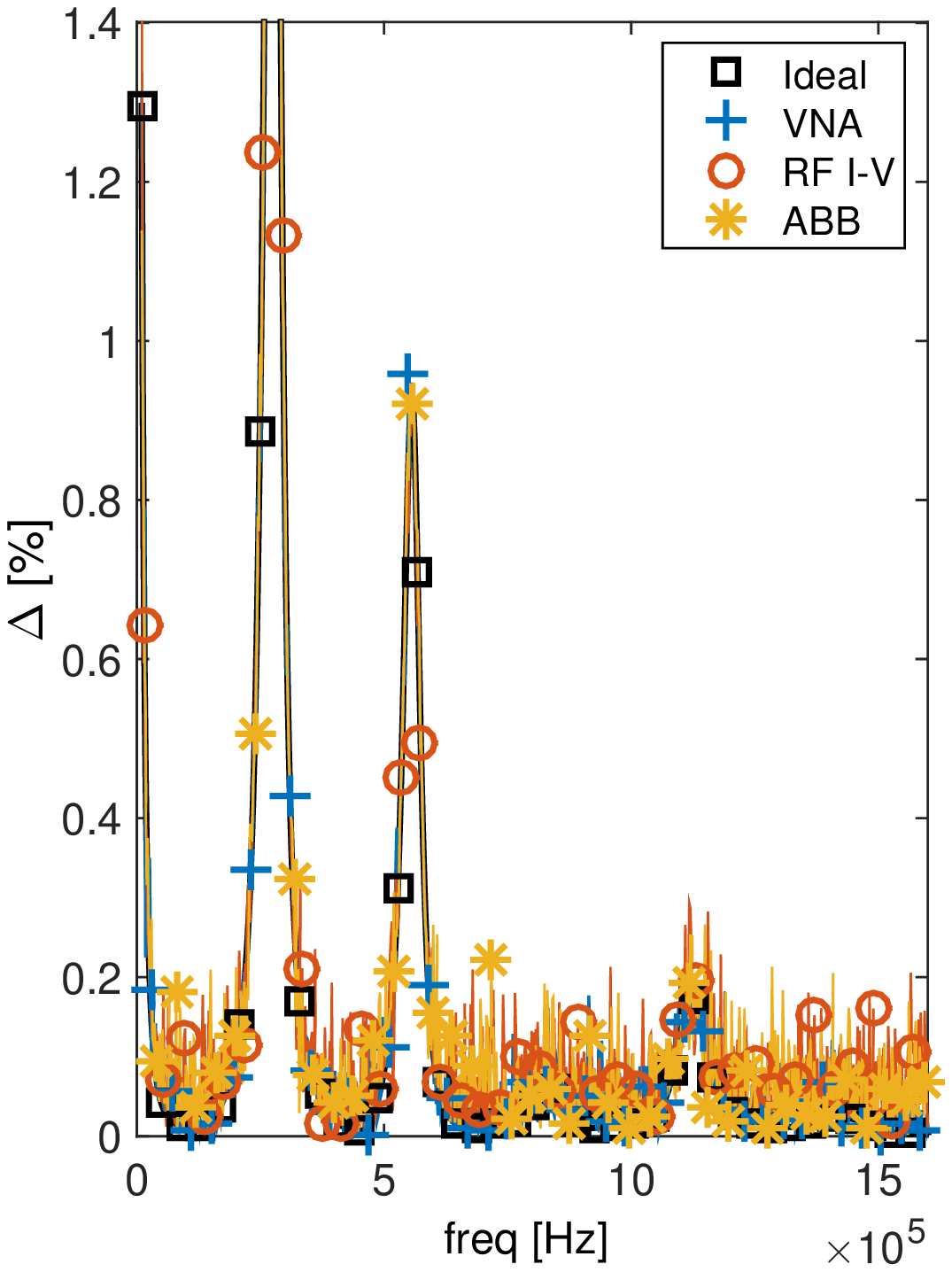}}\subfloat[$\rho$ based method]{\centering{}\includegraphics[width=0.5\columnwidth]{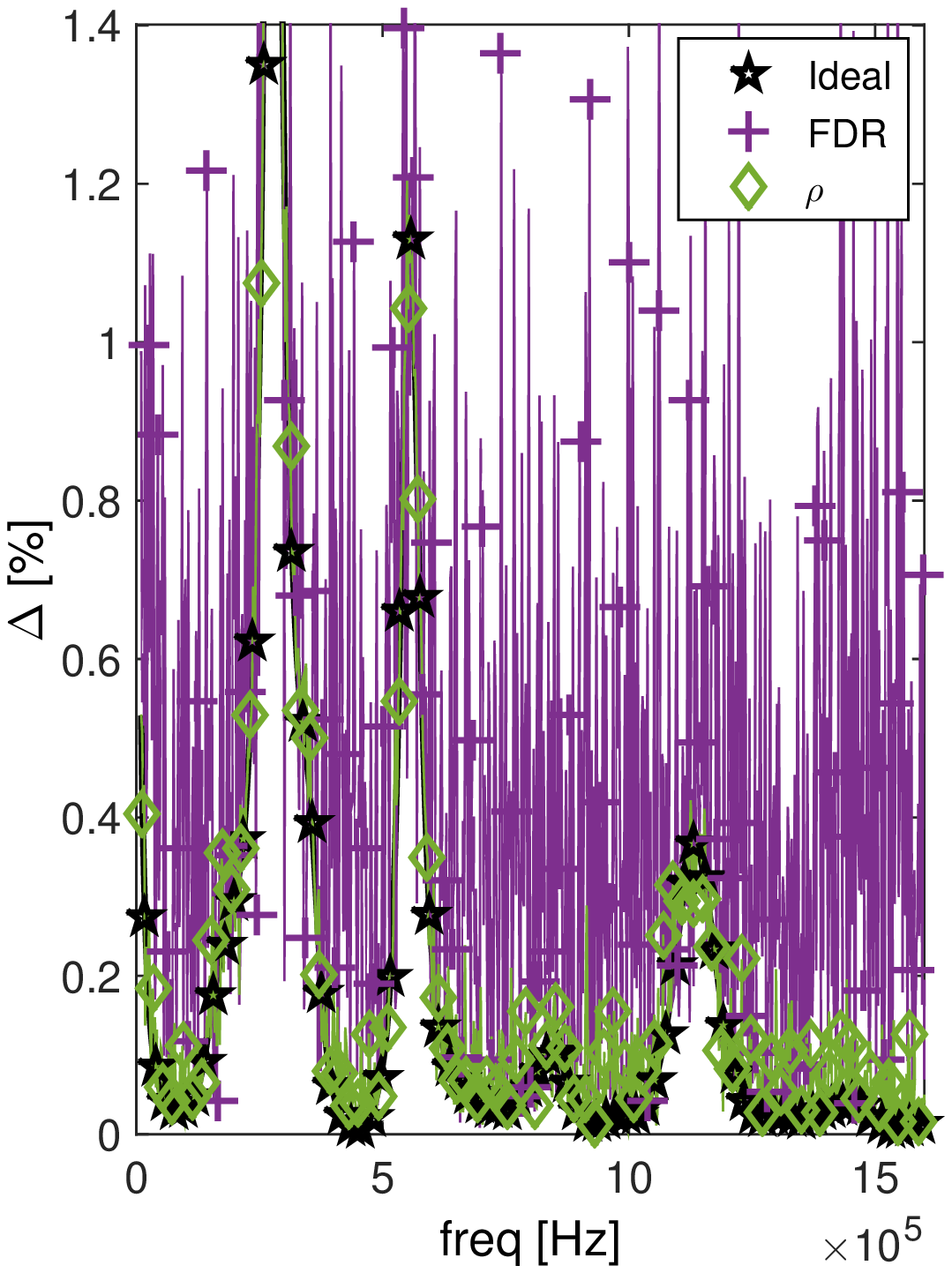}}\caption{$\Delta$ generated by either a high impedance or a very far fault.\label{fig:deltaf}}
\end{figure}
\begin{figure}[tb]
\centering{}\subfloat[Impedance based method]{\centering{}\includegraphics[width=0.5\columnwidth]{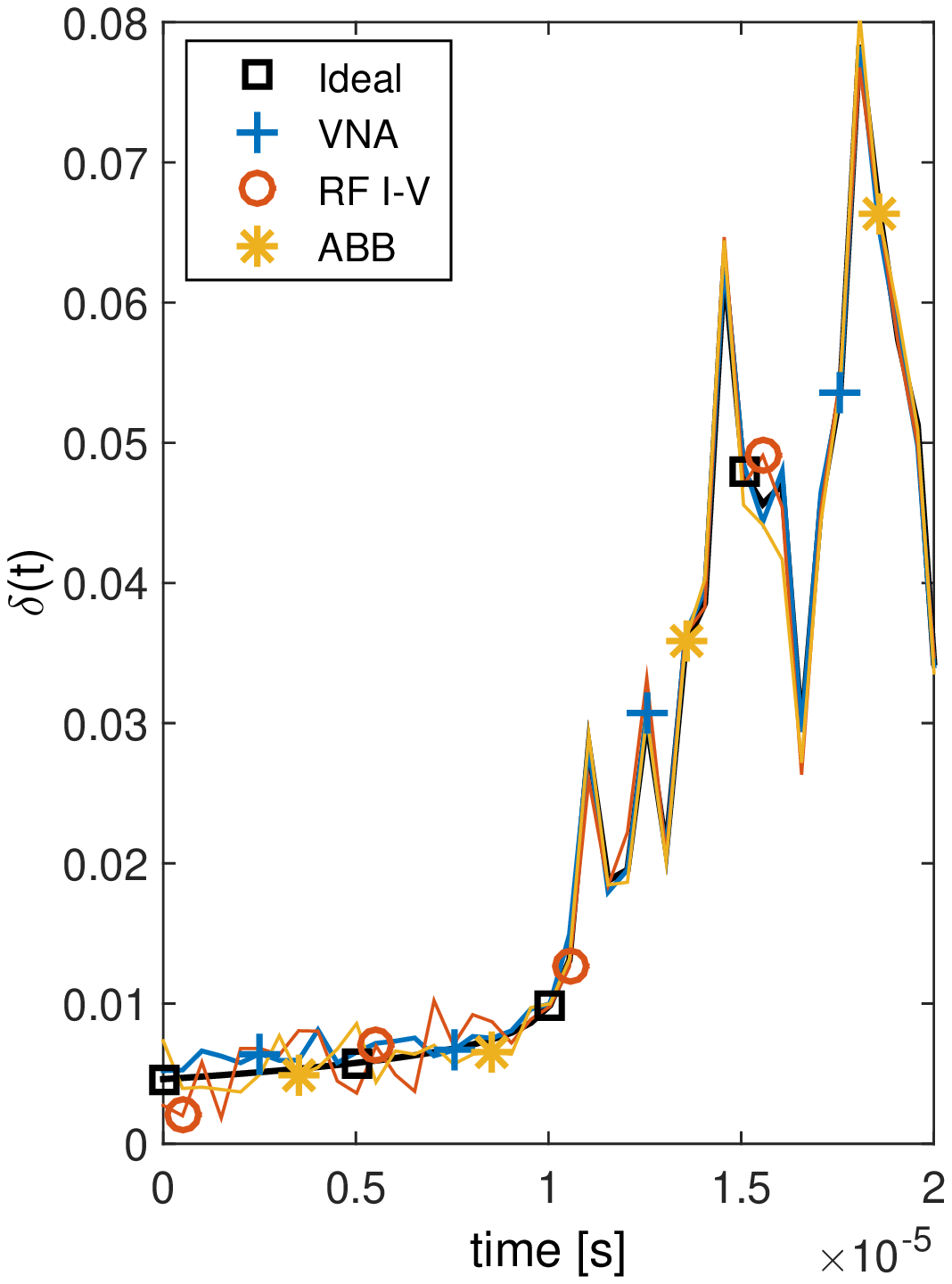}}\subfloat[$\rho$ based method]{\centering{}\includegraphics[width=0.5\columnwidth]{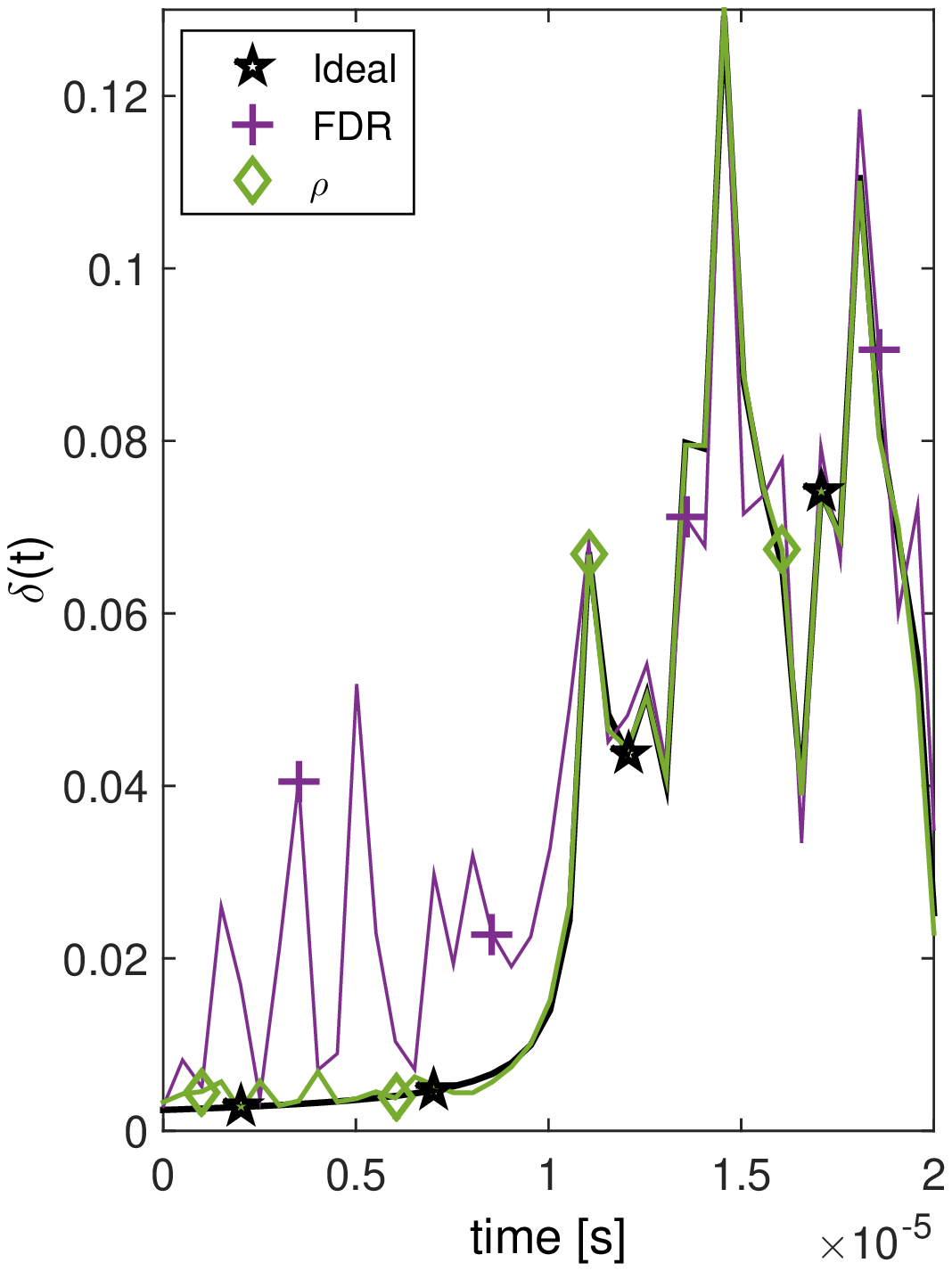}}\caption{$\delta(t)$ generated by either a high impedance or a very far fault.\label{fig:deltat}}
\end{figure}

As presented in \cite{7897102}, an electric fault can be detected
in power grids by continuously sensing the impedance over a wide spectrum.
More in general, a fault can be detected by continuously sensing the
network by means of any reflectometric method and then analyzing the
function
\[
\Delta=100\cdot\left|\frac{X_{f}-X_{p}}{X_{p}}\right|,
\]
 where $X$ can be either $Z_{PL}$ , $\rho$, or $T$ and the subscripts
$f$ and $p$ denote the fault and pre-fault conditions. If $\Delta$
is greater than a certain threshold defined by the user at a certain
frequency, then its inverse Fourier transform (IFT) $\delta(t)$ can
give useful information about the fault properties. In particular,
the first peak of $\delta(t)$ indicates the return time of the echo
caused by the fault. If the propagation velocity inside the network
is known, then the return time can be converted in the distance of
the fault.

In order to test the different measurement techniques presented in
this paper, we simulated the presence of a fault whose occurrence
results in low values of $\Delta$. Such a fault can be either characterized
by a very high impedance or simply occur far away from the measurement
point. The test signal and the power line noise are set at -55 dBm
and -120 dBm respectively. Looking first at the ideal measurements
in Fig. \ref{fig:deltaf}, we note that the values of $\Delta$ given
by an impedance measurement are in general lower compared to those
given by a measurement of $\rho$. This is due to the fact that, while
$\rho$ can linearly track changes of impedance along the network,
$Z_{PL}$ undergoes lower variation when very low or very high impedances
slightly change (see Fig. \ref{fig:Relation-between-}). Moving to
the noisy impedance measurements, we see that while the PLM-VNA method
almost corresponds to the ideal trace, the PLM-IV and PLM-ABB methods
give more noisy results. Similar results are given by the measurement
of $\rho$ using the PLM-VNA, while the performance of the FDR method
is significantly worse. This is due to the different level of noise
saturation reached by every method (see Fig. \ref{fig:INR-and-TNR}).
The analysis of $\delta(t)$ reveals some further insight: while the
impedance measured with the PLM-VNA method almost perfectly coresponds
to the ideal trace, the PLM-IV and PLM-ABB methods are characterised
by a higher level of noise in the first part and eventually fail to
identify a secondary peak. The noise of the $\rho$ measurement in
the first part is comparable to that of the PLM-ABB method, while
the noise of the FDR method is so high that a correct identification
of the first peak is impossible.

As expected from the preliminary analysis of the previous subsections,
the impedance measured with the PLM-VNA method also gives the best
results when applied to the fault detection problem. The PLM-ABB,
PLM-IV, and $\rho$ methods provide similar results in terms of
noise, but the $\rho$ measurement provides higher values of $\Delta$
resulting in easier fault detection. Finally, the FDR method provides
almost only noise data for such low values of $\Delta$.

\section{Conclusions\label{sec:Conclusions}}

In this paper, we proposed to integrate an additive circuit in the
front-end of power line communications modems, which allows them to
act as high frequency network sensors. We took under analysis three
different circuit architectures: the PLM-VNA, the PLM-ABB and the
PLM-IV. They enable to measure three quantities: impedance, reflection
coefficient and FDR trace. In order to assess the accuracy of such
measurements, we analyzed the effect of both the power line channel
noise and the PLM noise. The results showed that for low channel impedance
values the best accuracy in terms of QNR is obtained by the impedance
measured with the PLM-ABB architecture, while the one measured with
the PLM-VNA architecture is the best elsewhere. In this context, we
showed the values of QNR that are expected for common levels
of transmitted signal power and noise in both narrowband
and broadband PLC. In particular, we found the limits for which the
QNR is dominated by the power line noise and by the PLM noise respectively.
We also showed that, when applied to PL in the presence of background
noise, all the presented techniques provide measurements perturbed
by additive Gaussian noise. 

Finally, we applied the different measurement techniques to the problem
of fault detection. This application confirms the best performance
of the impedance measured with the PLM-VNA architecture, and also
points out that the widely used FDR technique gives the worse results
among the other considered techniques. In conclusion, the choice of
the architecture to be integrated in a PLM and the type of measurement
to be performed is based on the kind of network that has to be sensed.
For low voltage distribution grids, where the impedance has often
very low values, we suggest to perform network sensing by measuring
the network impedance with either the PLM-ABB or the PLM-IV architectures;
for medium voltage distribution grids or indoor grids, where the impedance
ranges from tents to hundreds $\Omega$, we suggest to measure the network
impedance with the PLM-VNA architecture.

\appendices{}

\section*{Appendix \label{sec:Appendix-A}}

Since in this paper all the formulas are frequency dependent and we
want to compare different measurement methods at each frequency, the
noise has to be defined in frequency domain. From the stochastic processes
theory, the Fourier transform $X(f)$ of a stationary process $x(t)$
with autocorrelation $R(t_{1},t_{2})=R(\tau)$ where $\tau=t_{1}-t_{2}$,
and power spectrum $S(\omega)$, is a random process with the following
properties \cite{papoulis2002}:
\begin{enumerate}
\item The mean $m(f)$ of $X(f)$ is the Fourier transform of the mean $m(t)=m_{0}$
of $x(t)$, namely
\begin{equation}
m(f)=m_{0}\delta(f)
\end{equation}

\item The autocorrelation $\Gamma(f_{1},f_{2})$ of $X(f)$ is
\begin{equation}
\Gamma(f_{1},f_{2})=2\pi S(f_{1})\delta(f_{1}-f_{2})\label{eq:fr_noise_corr}
\end{equation}

\end{enumerate}
hence the Fourier transform of a stationary process is a non-stationary
white process with spectral density $2\pi S(f_{1})$. Moreover, since
we are considering Gaussian noise processes, their Fourier transform
are still Gaussian processes. Therefore, using also \eqref{eq:fr_noise_corr},
the noise at each frequency $f_{i}$ is uncorrelated to the other
frequencies, and it is a complex Gaussian variable with variance $\sigma_{N}^{2}=2\pi S(f_{i})$.
Since the average power spectrum is known from different measurement
campaigns (see Section \ref{sec:Considerations-on-the}), the complex
Gaussian noise can be generated for each frequency using for example
the Box-Muller method \cite{box1958}. 

When the noise is generated in frequency domain, any colored power
spectral density can be accurately reproduced, and the relative noise
in time domain can be simply obtained by performing the inverse Fourier
transform of the generated noise. If we now consider, as it happens
in real systems, that the signal is transmitted at discrete frequencies,
the noise process generated with this method becomes continuous periodic
in time domain. In order to avoid time aliasing and the consequent
power spectral density alteration, the frequency step has to be short
enough to let $R(\tau)$ reach negligible values before the following
periodic repetition.

\bibliographystyle{IEEEtran}
\bibliography{biblio}

\end{document}